\documentclass[showpacs,prb,aps,superscriptaddress,twocolumn,floatfix]{revtex4-1}

\usepackage{amsmath}
\usepackage{amssymb}
\usepackage{graphicx}
\usepackage{bm}

\newcommand{\eps}{\varepsilon}
\newcommand{\sss}{\scriptscriptstyle}
\newcommand{\la}{\lambda}

\newcommand{\om}{\omega}
\newcommand{\K}{\mathrm{K}}

\newcommand{\prt}{\partial}

\allowdisplaybreaks[4]

\begin{document}

\title{Wave pattern induced by a localized obstacle
in the flow of a one-dimensional polariton
  condensate}

\author{P.-\'E. Larr\'e}\affiliation{Univ. Paris Sud, CNRS, Laboratoire de
  Physique Th\'eorique et Mod\`eles Statistiques, UMR8626, F-91405
  Orsay, France}
\author{N. Pavloff}\affiliation{Univ. Paris Sud, CNRS, Laboratoire de
  Physique Th\'eorique et Mod\`eles Statistiques, UMR8626, F-91405
 Orsay, France}
\author{A. M. Kamchatnov}\affiliation{Institute of Spectroscopy,
  Russian Academy of Sciences, Troitsk, Moscow Region, 142190, Russia}

\date{\today}

\begin{abstract}
  Motivated by recent experiments on generation of wave patterns by a
  polariton condensate incident on a localized obstacle, we study the
  characteristics of such flows under the condition that irreversible
  processes play a crucial role in the system. The dynamics of a
  non-resonantly pumped polariton condensate in a
  quasi-one-dimensional quantum wire is modeled by a Gross--Pitaevskii
  equation with additional phenomenological terms accounting for the
  dissipation and pumping processes. The response of the condensate
  flow to an external potential describing a localized obstacle is
  considered in the weak-perturbation limit and also in the nonlinear
  regime. The transition from a viscous drag to a regime of wave
  resistance is identified and studied in detail.
\end{abstract}


\maketitle

\section{Introduction}

The ability to move with respect to an obstacle without dissipating
energy is one of the most intuitive and appealing definition of
superfluidity. This is the reason why the motion of quantum fluids
with respect to obstacles has been used in several experiments aiming
at revealing a superfluid behavior in different physical systems:
$^4$He (see, e.g., Refs.~\onlinecite{All76,Ave85}), $^3$He (Ref.~
\onlinecite{Cas86}), ultracold atomic
vapors \cite{Ram99,Ono00,Mil07,Eng07,Dri10} and more recently
polariton condensates \cite{Amo09a,Amo09b,Nar11,Amo11,San11}.

For a weakly perturbing impurity moving at constant velocity $V$ in
a conservative atomic Bose--Einstein condensed (BEC) system at zero
temperature, the Landau criterion \cite{landau} predicts that there
exists a critical velocity $V_{\rm crit}$ separating two different
behaviors: (i) for $V<V_{\rm crit}$ no excitations are emitted away
from the obstacle and, hence, there is no drag force; (ii) for
$V>V_{\rm crit}$ a Cherenkov radiation of linear waves occurs; these
waves carry momentum away from the impurity which is thus subject to a
finite drag force. The first regime is superfluid and the second one
is dissipative\cite{remi}.

In a pumped non-equilibrium polariton condensate, even when
kinematically allowed, propagating disturbances are always damped due
to the finite lifetime of the polaritons. As a result, the well
defined transition between superfluid and dissipative regimes
transforms in these damped systems into a crossover characterized by
different forms of wave patterns: localized for small enough flow
velocity; oscillatory and extended for large enough flow velocity. The
boundary between these two regimes is typically not abrupt: just at
the transition point the decay length of a propagating wave is less than its
wavelength and this disturbance can hardly be distinguished from a
localized perturbation. It might thus be difficult to separate a
superfluid regime from a dissipative one by studying the wave pattern
created by an obstacle. Nevertheless, the concept of superfluidity
is often employed because it permits a simple qualitative discussion
of the processes taking place in the flow of a polariton condensate.

In the present work we study in detail the wake of a polariton
condensate past an obstacle and the associated drag force. We argue
that, for low enough damping, the superfluid/dissipative transition is
better understood in term of a crossover of the force experienced by
the obstacle from a viscous drag to wave resistance, in analogy
to what is observed for capillarity-gravity waves.

The paper is organized as follows. In Sec.~\ref{model} we present the
phenomenological one-dimensional model we use and present our strategy
for studying the specific features of typical flows. In
Sec.~\ref{perturbation} we set up a general perturbative analysis of
the motion of the polariton gas past a weak obstacle and discuss the
domain of validity of this approach. In Sec.~\ref{nonlinear} we obtain
non-perturbative results valid for a localized narrow impurity using
several approximation schemes (the so-called hydraulic approximation
in Sec.~\ref{hydraulic} and Whitham averaging method in
Sec.~\ref{whitham}) and also numerical integration
(Sec.~\ref{numerics}). Finally we present our conclusions in
Sec.~\ref{conclusion}. Some technical points are given in the
appendices. In Appendix \ref{AppA} we study the poles of the response
function of the system and in Appendix \ref{AppC} we present the
Whitham theory we use in Sec.~\ref{whitham} of the main text.

\section{The model}\label{model}

We study the flow of a polariton condensate past an obstacle
disregarding possible effects of polarization of the light modes in the
cavity. We consider a configuration in which excitons are confined in
a one-dimensional quantum wire and, as a result, the polariton
condensate is described by an order parameter $\psi(x,t)$ whose
dynamics is modeled by a Gross--Pitaevskii equation of the form
\begin{equation}\label{eq:scal1}
{\rm i}\hbar\,\psi_{t}=-\frac{\hbar^{2}}{2m}\, \psi_{xx}+
\big(U_{\mathrm{ext}}(x,t)+\alpha\rho\big)\psi+
{\rm i} \big(\gamma-\Gamma\rho\big)\psi.
\end{equation}
In Eq.~\eqref{eq:scal1} $m$ is the polariton effective mass (in the
parabolic dispersion approximation, valid at small momenta),
$\rho(x,t)=|\psi(x,t)|^2$ is the polariton density and $U_{\rm
  ext}(x,t)$ describes the potential of a loca\-li\-zed obstacle,
possibly in motion relative to the polariton gas. Interaction effects
are described by an effective local repulsive term characterized by
the nonlinear coupling constant $\alpha>0$. There is a whole body of
evidence showing that the overall effective interaction between
polaritons is repulsive. Some of the most direct manifestations of
this repulsion are the observed emission blueshift
\cite{Kas06,Baj08,Uts08} and the expulsion of the condensate from a
pumping region \cite{Wer06,Fer10}. Another consequence of repulsion,
very important for the present study, is the absence of scattering
from a defect---first observed in Refs.~\onlinecite{Amo09a,Amo09b}---and
the related emission of nonlinear excitations\cite{remi} (solitons and
vortices) whose generation is typically associated to a loss of
superfluidity \cite{Nar11,San11,Amo11,Gro11}.

Due to the finite lifetime of the polaritons, the system needs to be
pumped. Following Refs.~\onlinecite{Wou07,Kee08,Wou08,Wou10}, we
schematically describe this effect by the last term of
Eq.~\eqref{eq:scal1}: the term $\hbar\,\psi_{t}=\gamma\,\psi$
phenomenologically describes the combined effects of the pumping and
decay processes and for $\gamma>0$ an overall gain leads, if not
compensated, to an exponential increase of the density. This increase
is counterbalanced by the term $\hbar\,\psi_{t}=-\Gamma\rho\,\psi$
(where $\Gamma>0$) which accounts for a saturation of the gain at
large density and allows to reach a steady state
configuration---resulting from dynamical equilibrium between gain and
losses---with a finite density $\rho_0=\gamma/\Gamma$.
Eq.~\eqref{eq:scal1} corresponds to a situation where the pumping
extends over all space. This models a system where an obstacle is
present within a large reservoir, and simplifies the theoretical
treatment because the stationary density in absence of external
potential is constant. Results where the obstacle is present outside
of the pumping region will be presented in a forthcoming
publication\cite{nous}.

Localized structural defects are naturally present in many samples;
they can also be artificially created by means of lithographic
techniques or by a continuous-wave laser. If an obstacle
is introduced into the condensate, the state with uniform density
$\rho_0$ is disturbed. We suppose that the obstacle is described by a
potential $U_{\rm ext}(x,t)$ with a finite spatial extension
[verifying $U_{\rm ext}(x,t)\to 0$ as $|x|\to\infty$]. In many
experiments the condensate is put into motion with respect to the
obstacle by resonant pumping. Here we rather describe a situation with
non-resonant pumping, where condensation can be forced to occur in a
finite-momentum state by seeding the system with a short
coherent-light pulse \cite{Wou10}. However, we believe that the gross
features of the theoretical study of the wave patterns and of the drag
force are not essentially affected by the technique used for setting
the fluid into motion. This is supported by a comparison of the
results of the present work with the one of Ref.~\onlinecite{Ber12}
where a continuous transition at a critical velocity (possibly
different from the speed of sound) is also observed in a perturbative
study of a resonantly driven polariton fluid.

As just discussed, in typical experiments with polariton condensates
the obstacle does not move and instead the condensate is put into
motion with some velocity $V$. However we shall sometimes use for
convenience a reference frame in which the condensate is at rest (far
enough from the obstacle) and where the obstacle moves with velocity
$-V$: $U_{\rm ext}(x,t)=f_{\rm ext}(x+Vt)$.
A comprehensive study of this problem can be done in the case of an
obstacle represented by a weak potential which induces a wave
disturbance corresponding to small modifications of the parameters of
the flow. In this configuration the problem can be treated in the
framework of perturbation theory which is presented in the next
section of the paper. This corresponds to the extension to damped
systems of previous perturbative studies of BEC atomic 
vapors~\cite{Kov01,lp-2001,pavloff02,Ast04}. In this approach the case 
of a $\delta$-peak impurity is of special interest because its solution
corresponds to the Green function of the problem, and we treat it with
special care. In Sec.~\ref{nonlinear} instead, we consider the wave
pattern generated by the flow of a polariton condensate past a {\it
  strong} obstacle potential, when perturbation theory does not longer
apply. In this case, it is appropriate to distinguish between {\it
  wide} and {\it narrow} obstacles depending on the ratio of their
sizes to the healing length $\xi$ ($\xi$ is the de Broglie wavelength
of polaritons moving with the sound velocity; see its definition in
the next paragraph). When a narrow obstacle moves at supersonic speed the
downstream profile has a rather smooth behavior which can be described
by a dispersionless approach, the hydraulic approximation which we present
in Sec.~\ref{hydraulic}. On the other hand, the upstream-wave
structure can be represented (for small enough damping coefficient) as
a weakly modulated nonlinear periodic wave which is a damped
dispersive shock wave. Such shocks have been
studied for the case of a wide obstacle with the use of Whitham
modulation theory in Ref.~\onlinecite{Kam12}. In the present work we
present a similar and more detailed study in the case of a $\delta$-impurity in
Sec.~\ref{whitham}.

In absence of external potential, a homogeneous and stationary solution
of Eq.~\eqref{eq:scal1} corresponds to an order parameter of the form
$\psi(x,t)=\sqrt{\rho_0}\exp(-{\rm i} \mu t /\hbar)$, where $\rho_0$ is the
uniform density and $\mu$ is the chemical potential. Inserting this
expression in \eqref{eq:scal1} one finds $\rho_0=\gamma/\Gamma$
(necessary for obtaining a real $\mu$ corresponding to a time-
independent density) and $\mu=\alpha\rho_0$. The characteristic
density $\rho_0$ and energy $\mu$ are associated to characteristic
velocity and distance, namely the speed of sound\cite{remcs}
$c_s=\sqrt{\alpha\rho_0/m}$ and the healing length
$\xi=\hbar/(mc_s)$.

We will see below that, for a given
obstacle potential $U_{\rm ext}(x,t)$, the flow pattern is monitored by
only two dimensionless parameters: the Mach number $M$ and the
damping parameter $\eta$ defined as
\begin{equation}\label{eq:meta}
M=\frac{V}{c_s}\quad\textrm{and}\quad\eta=\frac{\gamma}{\mu}.
\end{equation}
Having identified the relevant parameters of the problem one can
simplify the notations by expressing densities in units of $\rho_0$,
distances in units of $\xi$, times in units of $\xi/c_s$ and energies
in units of $\mu$. In these new variables Eq.~(\ref{eq:scal1}) takes
the form
\begin{equation}\label{eq:4-1}
{\rm i}\,\psi_{t}=-\tfrac{1}{2}\psi_{xx}+\big(U_{\mathrm{ext}}(x,t)+
\rho\big)\psi+{\rm i}\eta\big(1-\rho\big)\psi.
\end{equation}
From now on, we shall use this dimensionless form of the
damped Gross--Pitaevskii equation.

\section{Flow past a weak obstacle}\label{perturbation}

\subsection{General linear theory}\label{3a}

In absence of external potential Eq.~\eqref{eq:4-1} admits a uniform
stationary solution of the form $\psi(x,t)=\exp(- {\rm i} \, t)$. If the
potential of the obstacle is weak, one can evaluate the density and
the flow velocity profiles of the polariton condensate
perturbatively. In this case one looks for a solution of
Eq.~\eqref{eq:4-1} of the form
\begin{equation}\label{eq:scal4}
\psi(x,t)=
\left[1 +\varphi(x,t)\right] \exp(- {\rm i} \, t),
\end{equation}
assuming that $|\varphi(x,t)| \ll 1$. Linearizing Eq.~\eqref{eq:4-1}
with res\-pect to $\varphi(x,t)$ and $U_{\mathrm{ext}}(x,t)$ and
introducing the Fourier transforms
\begin{equation}
\label{eq:Fourier transforms}
\begin{bmatrix}
u(q,\omega) \\ v(q,\omega) \\ \hat{U}_{\mathrm{ext}}(q,\omega)
\end{bmatrix}
=\int_{\mathbb{R}^{2}}\mathrm{d}x\,\mathrm{d}t
\begin{bmatrix}
\varphi(x,t) \\ \varphi^{\ast}(x,t) \\ U_{\mathrm{ext}}(x,t)
\end{bmatrix}
{\rm e}^{-{\rm i}(qx-\omega t)},
\end{equation}
one finds that $u(q,\omega)$ and $v(q,\omega)$ satisfy the following
linear system:
\begin{equation}
\label{eq:Linear system verified by u and v}
\mathcal{L}
\begin{pmatrix}
u(q,\omega) \\ v(q,\omega)
\end{pmatrix}
=-\hat{U}_{\mathrm{ext}}(q,\omega)
\begin{pmatrix}1 \\ 1\end{pmatrix},
\end{equation}
where
\begin{equation}
\label{eq:Matrix L}
\mathcal{L}=
\begin{pmatrix}
\frac{q^{2}}{2}-\omega+1-{\rm i}\eta &
1-{\rm i}\eta \\
1+{\rm i}\eta &
\frac{q^{2}}{2}+\omega+1+{\rm i}\eta
\end{pmatrix}.
\end{equation}
When $\hat{U}_{\mathrm{ext}}(q,\omega)\equiv0$, i.e., in the absence
of the external obstacle, non-trivial solutions $u(q,\omega)$ and
$v(q,\omega)$ of the $2\times2$ system \eqref{eq:Linear system verified
  by u and v} exist only when the determinant
\begin{equation}
\label{eq:Determinant of the linear system}
D(q,\omega)=
q^{2}\left(1+\frac{q^{2}}{4}\right)-\omega^{2}-
2{\rm i}\eta\,\omega
\end{equation}
of the matrix $\mathcal{L}$ is identically null. The resolution of the
characteristic equation $D(q,\omega)=0$ yields the dispersion relation
$\omega(q)$ of the elementary excitations propagating on top of a
homogeneous and stationary profile. Let us first consider the case
$\eta\to 0$ (and also in dimensional units $\Gamma\to 0$ in such a way
that the density $\rho_0=\gamma/\Gamma$ is kept constant). In this case one
finds that the excitation spectrum is the Bogoliubov one, i.e., one recovers the dispersion
relation of elementary excitations of a weakly interacting atomic Bose gas:
$\omega(q)=\pm\,\omega_{\rm B}(q)$, where
\begin{equation}
\label{eq:Bogoliubov spectrum}
\omega_{\rm B}(q)=
q\,\sqrt{1+\frac{q^{2}}{4}}.
\end{equation}
In the case where $\eta$ is not zero one
gets \cite{Wou07}
\begin{equation}
\label{eq:Dispersion relation}
\omega(q)=\left\{
\begin{array}{lcl}
-{\rm i}\eta \pm
{\rm i}\sqrt{\eta^{2}-\omega_{\rm B}^{2}(q)}
& \mbox{if} & |q|<q_{\ast},
\vspace{3mm} \\
-{\rm i}\eta
\pm
\sqrt{\omega_{\rm B}^{2}(q)-\eta^{2}}
& \mbox{if} & |q|>q_{\ast},
\end{array}
\right.
\end{equation}
where
\begin{equation}
\label{eq:qstar}
q_{\ast}=\left[2\left(\sqrt{1+\eta^{2}}-1\right)\right]^{1/2}.
\end{equation}
In the ideal case ($\eta=0$ and then $q_*=0$) long-wavelength perturbations
($|q|\ll 1$) correspond to sound waves with a linear dispersion $\omega_{\rm
  B}(q)\simeq q$ and with a sound velocity equal to
unity in our dimensionless units. As announced in note
\onlinecite{remcs}, perturbations with $|q|<q^*$ do not propagate in
presence of finite damping ($\eta\ne 0$). However, for small $\eta$
there exists a finite region of wavenumber ($q_*\ll |q| \ll 1$)
for which the dispersion relation \eqref{eq:Dispersion relation} can
be approximated by the long-wavelength limit $\omega(q)\simeq q-{\rm i}\eta $
describing weakly damped sound-waves.

Let us now consider the general case
where $U_{\rm ext}(x,t)$ is not zero: the linear waves are generated by
the external potential and their Fourier components $u(q,\om)$ and
$v(q,\om)$ can be expressed by means of Eq.~(\ref{eq:Linear system
  verified by u and v}) in terms of this potential. This yields the
following expression for the first order density modulation
$\delta\rho=\varphi+\varphi^*$ induced by $U_{\rm ext}(x,t)$:
\begin{equation}
\label{eq:Dfluc}
\delta\rho(x,t)=
\int_{\mathbb{R}^{2}}
\frac{\mathrm{d}q\,\mathrm{d}\omega}{(2\pi)^{2}}\,
\chi(q,\omega)\,\hat{U}_{\mathrm{ext}}(q,\omega)\,
{\rm e}^{{\rm i}(qx-\omega t)},
\end{equation}
where
\begin{equation}
\label{eq:Linear-response-function}
\chi(q,\omega)\equiv
\frac{\delta\hat{\rho}(q,\omega)}{\hat{U}_{\mathrm{ext}}(q,\omega)}
=- \frac{q^2}{D(q,\omega)}
\end{equation}
is the linear response function of the system.
A configuration of great experimental interest corresponds to the case where
the condensate moves at constant velocity with respect to a static
obstacle. In this case, in the frame where the condensate is at rest,
the external potential is of the form
\begin{equation}
\label{eq:movpot}
U_{\mathrm{ext}}(x,t)=f_{\mathrm{ext}}(x+Mt),
\end{equation}
where $M$ is, in our dimensionless units, the velocity of the obstacle
with respect to the condensate. For being specific, we shall
henceforth consider the case $M>0$ which corresponds to an obstacle
moving to the left in a frame where the condensate is at rest.
Denoting by $\hat{f}_{\rm ext}$ the Fourier transform of $f_{\rm ext}$
[i.e., $\hat{f}_{\mathrm{ext}}(q) = \int_{\mathbb{R}}\mathrm{d}z\,
f_{\mathrm{ext}}(z)\exp(-{\rm i} q z)$] the expression of
$\delta\rho(x,t)$ in the case of an external potential of the form
\eqref{eq:movpot} reads
\begin{equation}\label{eq:generic}
\begin{split}
\delta\rho(x,t)&=\int_{\mathbb{R}}\frac{\mathrm{d}q}{2\pi}\,
\chi(q,-Mq)\,\hat{f}_{\mathrm{ext}}(q)\,
{\rm e}^{{\rm i}q(x+M t)} \\
&= \int_{\mathbb{R}}{\rm d}z \, K(x+M t-z)\, f_{\mathrm{ext}}(z),
\end{split}
\end{equation}
where
\begin{equation}\label{eq:K}
K(X)=\int_{\mathbb{R}}\frac{\mathrm{d}q}{2\pi}\,
\chi(q,-Mq) \,
{\rm e}^{{\rm i} q X}.
\end{equation}
One can first remark that $\delta\rho$ is a function of $x+Mt$ only:
the perturbative approach predicts that the density modulations
induced by an obstacle moving at constant velocity are stationary in
the reference frame where the obstacle is at rest.  Note however that,
in absence of damping, experiment performed on atomic condensates
\cite{Eng07} and theory \cite{lp-2001,pavloff02,legk} show that there
is a regime of time-dependent flows for impurity velocities close to
the speed of sound. This is a nonlinear effect which is missed by the
perturbative approach. In presence of damping this time-dependent
behavior also exits but, in a numerical study of nonlinear effects in
presence of a wide obstacle, it is observed in a smaller domain in the
parameter space $(\textrm{Intensity of }U_{\mathrm{ext}},V)$ than when $\eta\equiv 0$
\cite{Kam12}. This is confirmed in the case of a narrow obstacle by
the numerical results of Sec.~\ref{numerics} below. In this respect,
the perturbative result---being stationary---is thus more sound in
presence of damping since in this case the domain of time-dependent
flows is reduced. We make this discussion quantitative at the end of
Sec.~\ref{3b} by discussing the parameters governing the mathematical
validity of perturbation theory.

A particular property of solution \eqref{eq:generic} comes from
the conservation equation
\begin{equation}\label{eq:cons_mass}
\rho_t + j_x
=2\eta\,\rho\,(1-\rho),
\end{equation}
where $j=\mbox{Im}(\psi^*\psi_x)$ is the
particle current-density. Actually, Eq.~\eqref{eq:cons_mass} is a
{\it bona fide} conservation equation only when $\eta=0$. For nonzero
$\eta$, the number of particles is not conserved and Eq.~\eqref{eq:cons_mass} 
should rather be called a ``non-conservation'' equation
for the current of particles.
Eq.~\eqref{eq:cons_mass} is a direct consequence of
\eqref{eq:scal1}; in the stationary regime it implies \cite{Kee08,Kam12}
\begin{equation}\label{eq:cons_mass1}
\int_{\mathbb{R}}{\rm d}x \, \rho \, (1-\rho) = 0.
\end{equation}
At the perturbative level this reads $\int_{\mathbb{R}}{\rm
  d}x\, \delta\rho=0$ which is trivially verified by \eqref{eq:generic}.

\subsection{Flow past a $\delta$-impurity}\label{3b}

It is instructive to discuss in greater details the characteristics of
the wave pattern induced by a localized obstacle with the potential
\begin{equation}\label{eq:7-1}
U_{\rm ext}(x,t)=\varkappa\,\delta(x+Mt).
\end{equation}
Then one gets
\begin{equation}\label{7-2}
\delta\rho(x,t) = \varkappa \, K(X=x+Mt).
\end{equation}
This density modulation is typical for the perturbations induced by a
narrow obstacle moving in the polariton condensate. Besides, the
solution of the $\delta$-impurity problem is particularly interesting
because $K(X)$ is the Green function from which the result for any
potential is obtained by convolution [cf.~Eq.~\eqref{eq:generic}].

The integral \eqref{eq:K} can be computed by
the method of residues and $K(X)$ has different behaviors depending on
the value of $M$ and corresponding to different arrangements of the
poles of $\chi(q,-Mq)$ in the complex $q$-plane. The poles
are the roots of the equation $D(q,-Mq)/q=0$ which reads
\begin{equation}\label{7-2a}
q^3 + 4(1-M^2)q + 8{\rm i}\eta M=0.
\end{equation}
The explicit expression of the three poles $q_1$, $q_2$ and $q_3$ in function of $\eta$ and $M$ is
given in Appendix \ref{AppA}. One obtains the following generic expression
\begin{equation}\label{eq:Kx}
K(X) = {\rm i} \sum_{\ell=1}^3
\mbox{sgn}(\mbox{Im}\,q_\ell)\,\mbox{Res}(q_\ell) \,
\Theta[\mbox{sgn}(\mbox{Im}\,q_\ell)X] \,
{\rm e}^{{\rm i} q_\ell X},
\end{equation}
where $\Theta$ is the Heaviside step function and
$\mbox{Res}(q_\ell)$ is the residue of $\chi(q,-Mq)$ at $q_\ell$:
\begin{equation}\label{eq:res}
\mbox{Res}(q_\ell)=\frac{- 4 \, q_\ell}{3q_\ell^2+4(1-M^2)}.
\end{equation}

\begin{figure}
\includegraphics*[width=0.99\linewidth]{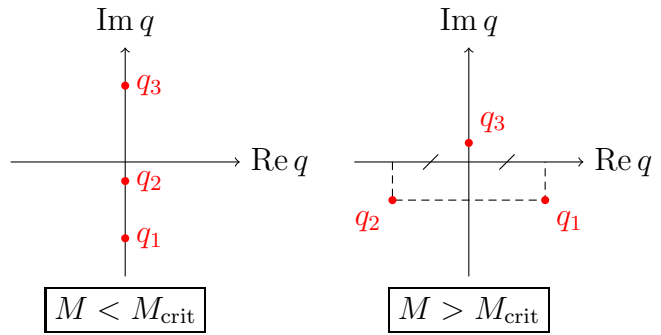}
\caption{(Color online) Location of the three poles $q_1$, $q_2$ and $q_3$
  of $\chi(q,-Mq)$ in the complex $q$-plane. For positive (negative)
  $X=x+Mt$ the integral in Eq.~\eqref{eq:K} is evaluated by closing
  the contour from above (below). As a result, for $M>M_{\rm crit}$
  (damped) density oscillations are observed upstream the obstacle
  (i.e., for $X<0$).}
\label{fig1}
\end{figure}

There exists a critical velocity $M_{\rm crit}$ below which the poles
of $\chi(q,-Mq)$ are all located on the imaginary axis
(cf.~Fig.~\ref{fig1} and also Appendix \ref{AppA}) and in this case
formula \eqref{eq:Kx} shows that $K(X)$ exponentially goes to
$0$ when $|X|\to\infty$. A more transparent expression can be
obtained by explicitly solving the third order equation
(\ref{7-2a}). This yields
\begin{equation}\label{eq:7-3}
\begin{split}
K(X\leqslant0)&=-\frac{2}{A}\bigg[\frac{A-B}{A-3B}\,{\rm e}^{(A-B)X} \\
& \hphantom{=-\frac{2}{A}\bigg[\frac{A-B}{A-3B}\,}
-\frac{4AB}{A^2-9B^2}\,{\rm e}^{2BX}\bigg], \\
K(X\geqslant0)&=-\frac{2}{A}\,\frac{A+B}{A+3B}\,{\rm e}^{-(A+B)X},
\end{split}
\end{equation}
where $A$ and $B$ are positive real numbers ($A>B\geqslant0$) depending on $M$
and $\eta$, whose expressions are given in Appendix \ref{AppA}
[Eq.~\eqref{eq:ap2-old}].

On the other hand, when $M>M_{\rm crit}$, two of the
poles acquire a real part and are symmetrically disposed with respect
to the imaginary axis (cf.~Fig.~\ref{fig1}). In this case
the wave pattern is given by the explicit formulas
\begin{equation}\label{eq:7-6}
\begin{split}
K(X\leqslant0) &=
-\frac{4}{E}\,
\mathrm{Im}
\left(\frac{E-{\rm i}F}{E-3{\rm i}F}\,{\rm e}^{{\rm i}EX}\right){\rm e}^{FX}, \\
K(X\geqslant0) &=
-\frac{8 F}{E^2+9F^2}\,{\rm e}^{-2FX},
\end{split}
\end{equation}
where the expression of the positive real numbers
$E$ and $F$ is given in Eq.~\eqref{eq:ap3old}.

The transition from one regime to the other takes place when two
roots of Eq.~(\ref{7-2a}) (namely $q_1$ and $q_2$) collide on the imaginary
axis, that is when the discriminant of this equation vanishes. This
condition yields the expression of $M_{\rm crit}$:
\begin{equation}\label{eq:vcr}
\begin{split}
M_{\rm crit}^2 = 1 - \frac{3}{2} \, \eta^{2/3}
\bigg[
&\Big(\sqrt{1+\eta^2}+1\Big)^{1/3} \\
- & \Big(\sqrt{1+\eta^2}-1\Big)^{1/3}
\bigg].
\end{split}
\end{equation}
When $\eta\to 0$, i.e., in the absence of damping, one reco\-vers the
usual Landau threshold for emission of Cherenkov radiation in a weakly
interacting Bose gas: $M_{\rm crit}=1$ (in dimensional units: $V_{\rm
  crit}=c_s$). In this case, the perturbative treatment states that
the flow is superfluid for velocities below $M_{\rm crit}$ and
dissipative above (see Refs.~\onlinecite{pavloff02,Ast04} and the
computation of the drag in Sec.~\ref{subdrag}). This is identical to Landau's
criterion since both approaches give the same value of velocity for
the onset of dissipation and have the same physical content:
excitation of small non-localized perturbations is allowed only above
$M_{\rm crit}$.

\begin{figure}
\includegraphics*[width=0.99\linewidth]{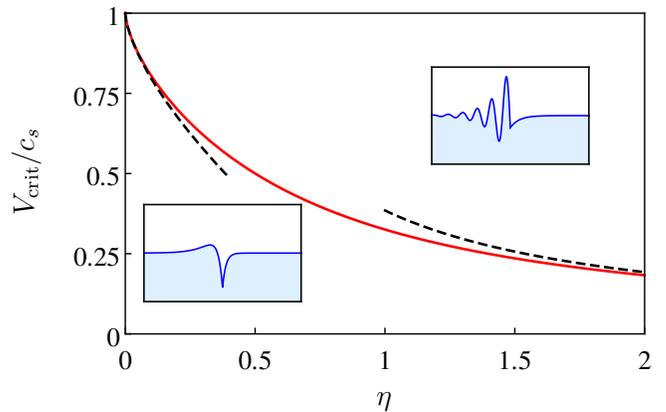}
\caption{(Color online) $M_{\rm crit}=V_{\rm crit}/c_s$ as a function
  of the dimensionless damping parameter $\eta$, such as given by
  Eq.~\eqref{eq:vcr}. The dashed lines correspond to the asymptotic
  expressions $M_{\rm crit}\simeq1-\frac{3}{2}(\eta/2)^{2/3}$ (for low
  $\eta$) and $M_{\rm crit}\simeq2/(3\sqrt{3}\,\eta)$ (for large
  $\eta$). The insets represent typical density profiles in presence
  of a repulsive $\delta$-peak impurity for $M<M_{\rm crit}$ (lower
  left inset) and $M>M_{\rm crit}$ (upper right inset).}
\label{fig2}
\end{figure}

In presence of dissipation $\eta\ne 0$, and Eq.~\eqref{eq:vcr} shows
that $M_{\rm crit}$ is a decreasing function of $\eta$
(cf.~Fig.~\ref{fig2}). For $M<M_{\rm crit}$ (subcritical velocities)
there is no Cherenkov radiation but, as shown by the explicit
computation of the drag force below, contrarily to the $\eta=0$ case,
the dissipative effects associated to the finite lifetime of
polaritons induce a finite drag force on the obstacle and the flow is
not superfluid. For $M>M_{\rm crit}$, Cherenkov radiation becomes
possible but dissipation within the condensate induces decay of the
associated density oscillations. The corresponding density patterns
are represented in each case ($M\lessgtr M_{\rm crit}$) in the insets
of Fig.~\ref{fig2} and the relevant analytical expressions are given
by Eqs.~\eqref{eq:7-3} and \eqref{eq:7-6}.

The fact that $M_{\rm crit}$ is modified by damping physically
explains why perturbation theory is more accurate in presence of
damping. For a non-damped system, an obstacle moving at velocity close
to $M_{\rm crit}=1$ generates Bogoliubov excitations whose typical
velocity is also close to $c_s=1$. As a result, the perturbations
accumulate in vicinity of the obstacle (since they propagate at the same 
velocity), nonlinear effects cannot be
neglected and the perturbative approach fails \cite{remflu}. In
presence of damping the critical velocity $M_{\rm crit}$ for radiating
Cherenkov waves differs from the velocity of propagation of small
amplitude perturbation and, moreover, the damping prevents large
increases of the density. As a result there is no pile up of
fluctuations in vicinity of the obstacle, nonlinear effects may be
neglected and the perturbative treatment is more likely to be valid.

This intuitive explanation of the increased accuracy of perturbation
theory in presence of damping is sustained by the mathematical
reasoning we present now.  In absence of damping the amplitude of the
relative density perturbation are of typical magnitude
$\varkappa/|M^2-1|^{1/2}$, i.e., perturbation theory indeed seriously
fails when the velocity of the obstacle is close to the speed of sound
\cite{lp-2001} because the expression for $\delta\rho$ diverges. This
problem is partially cured in presence of damping: for a potential of
the form \eqref{eq:7-1} a possible estimate of the amplitude of
$|\delta\rho(x,t)|$ is its value $\varkappa \, |K(0)|$ at the position
of the obstacle. A study of the dependence of this quantity on the
velocity and of the damping (i.e., on the dimensionless parameters $M$
and $\eta$) shows that, for a fixed value of $\eta$, it typically
reaches its largest value when $M=M_{\rm crit}$. The value of the
quantity $\varkappa \, |K(0)|$ at $M=M_{\rm crit}$ is thus the small
parameter $\epsilon$ of the perturbation expansion, in the sense that
if this quantity is small for given $\varkappa$ and $\eta$, the
perturbation theory is valid {\it for all velocities}. This condition
reads [see formula \eqref{eq:kde0}]
$\epsilon\equiv\varkappa/(1-M^2_{\rm crit})^{1/2}\ll 1$. Hence
$\epsilon$ is the small parameter of the perturbation theory in
presence of damping. It never diverges for finite $\eta$ and this
shows that perturbation theory is more sound with than without
damping. We see that $\epsilon$ effectively decreases in presence of
damping because $M_{\rm crit}$ differs from $1$, as advocated in the
intuitive discussion of the previous paragraph. For small $\eta$,
Eq.~\eqref{eq:ap10} yields $\epsilon\propto {\varkappa}\, \eta^{-1/3}$
whereas for large $\eta$ one finds $\epsilon\propto {\varkappa}$. One
can thus equivalently define the small parameter of the theory as
\begin{equation}\label{eq:ap10}
\epsilon=\varkappa\times\max\{1,\eta^{-1/3}\},
\end{equation}
and indeed a numerical check shows that, at fixed $\eta$, $\epsilon$
is a good estimate of the maximum value of $|\delta\rho(x)|$ for
$x\in\mathbb{R}$ and $M\in\mathbb{R}_+$.

We stress that the condition $\epsilon\ll 1$ is a criterion of
applicability of perturbation theory {\it for all $M$} at fixed $\eta$
and $\varkappa$. It is a strong requirement: for given $\eta$ and
$\varkappa$ failing to fulfill the condition $\epsilon\ll 1$, there
are still some velocities for which perturbation theory holds. For
instance in the supersonic regime, when $\eta M (M^2-1)^{-3/2}\ll 1$,
the condition of applicability of perturbation theory relies of the
smallness of the upstream oscillations and reads $\varkappa/
(M^2-1)^{1/2}\ll 1$.

\subsection{Generic flow pattern for a weak obstacle}\label{flowpattern}

For an obstacle of the generic form \eqref{eq:movpot} the position of
the poles of the response function and the critical velocity
\eqref{eq:vcr} play the same crucial role as for a $\delta$-impurity.
Eq.~\eqref{eq:generic} yields the following explicit expression for
the density oscillations:
\begin{equation}\label{eq:wake}
\begin{split}
\delta\rho(X) &=
 {\rm i} \int_{-\infty}^X {\rm d}y
\; \mbox{Res}(q_3) \, f_{\rm ext}(y) \, {\rm e}^{{\rm i} q_3(X-y)} \\
&-{\rm i} \int_{X}^{\infty} {\rm d}y
\sum_{\ell\in\{1,2\}}
\mbox{Res}(q_\ell) \, f_{\rm ext}(y) \, {\rm e}^{{\rm i}q_\ell(X-y)},
\end{split}
\end{equation}
where we recall that $\mbox{Res}(q_\ell)$ is the residue of $\chi(q,-Mq)$ at
$q_\ell$ ($\ell=1$, 2 or 3) [see Eq.~\eqref{eq:res}]. Formula
\eqref{eq:wake} is valid both below and above $M_{\rm crit}$. When
$\eta=0$ it reduces to the one already obtained in
Ref.~\onlinecite{lp-2001} in absence of damping [Eq.~(45) of this
reference].

It is interesting to obtain from \eqref{eq:wake} the generic form of
the long-distance wake which exists ahead of the obstacle when
$M>M_{\rm crit}$. When $X$ is negative and much larger than the range
of the obstacle potential $f_{\rm ext}$, the first term in
\eqref{eq:wake} can be neglected. If, furthermore, $f_{\rm ext}$
decreases rapidly enough at $-\infty$ so that $\hat{f}_{\rm
  ext}(q_{1,2})$ exists (typically when $f_{\rm ext}(x)$ decreases
more rapidly than $\exp[-\,{\rm Im}(q_{1,2}) \, x]$), one can
approximate the second integral by a compact expression yielding
\begin{equation}\label{eq:wake-long-dist}
\delta\rho(X)\underset{X\to-\infty}{\simeq}
2\,\mbox{Im}\left[
\mbox{Res}(q_1)\,\hat{f}_{\rm ext}(q_1)\,{\rm e}^{{\rm i} q_1 X}\right].
\end{equation}
We recall that Eq.~\eqref{eq:wake-long-dist} is an approximation of
formula \eqref{eq:wake} valid for $M>M_{\rm crit}$. It is of course
exact for all $X\leqslant 0$ in the case of a $\delta$-impurity. It
describes Cherenkov oscillations which are damped by a factor
$\exp[-\,{\rm Im}(q_1) \, x]$, in complete agreement with the results
obtained in Ref.~\onlinecite{Kam12} both numerically and also by means
of Whitham averaging method [Eq.~(42) of this reference].

Note that for large velocities ($M\gg M_{\rm
  crit}$) the imaginary parts of $q_1$ and $q_2$ tend to zero
(cf.~Appendix \ref{AppA}) and the wake \eqref{eq:wake-long-dist} thus
extends far ahead from the obstacle: the effective damping of the
Cherenkov radiation tends to zero. However, in this limit, $|q_1|$ gets
very large (cf.~Appendix \ref{AppA}) and for a generic potential
$|\hat{f}(q_1)|$ becomes very small: the amplitude of the wake decreases
uniformly at large velocity, not because of damping, but because the
large kinetic energy of the flow with respect to the obstacle allows
to treat this obstacle as a small perturbation. The same effect had
been predicted for BEC of ultracold vapors in
Ref.~\onlinecite{lp-2001} and has been observed experimentally in
Refs.~\onlinecite{Eng07,Dri10}.

\begin{figure}
\includegraphics*[width=0.99\linewidth]{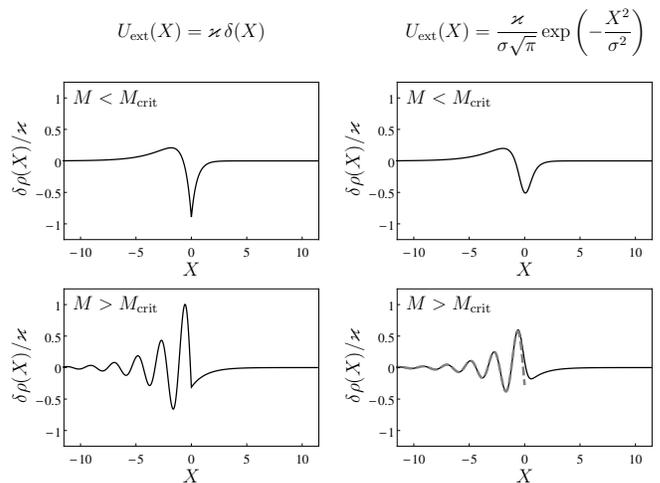}
\caption{$\delta\rho(X=x+Mt)$ for a $\delta$-impurity potential (left
  panels) and a Gaussian potential \eqref{eq:fw} of width $\sigma=0.5$
  (right panels) as given by perturbation theory [Eq.~\eqref{eq:generic}]. 
The plots are drawn for a system in which $\eta=0.5$ and
  in this case $M_{\rm crit}=0.5$. The two upper panels correspond to
  a velocity below $M_{\rm crit}$ ($M=0.4$) and the two lower ones to a
  velocity above $M_{\rm crit}$ ($M=1.75$). In the lower right panel the
  dashed gray line correspond to the approximation
  \eqref{eq:wake-long-dist}.}
\label{fig5}
\end{figure}

For being specific, we compare in Fig.~\ref{fig5} the density
modulations obtained within perturbation theory for a
$\delta$-impurity obstacle \eqref{eq:7-1} with the ones corresponding to a
Gaussian potential of finite width $\sigma$:
\begin{equation}\label{eq:fw}
U_{\rm ext}(x,t)=\frac{\varkappa}{\sigma\sqrt{\pi}}
\exp\left[-\frac{(x+Mt)^2}{\sigma^2}\right].
\end{equation}
When $\sigma\to 0$ this potential tends to the $\delta$-impurity
potential \eqref{eq:7-1}. As just explained, when $M>M_{\rm crit}$ the
damping of the oscillatory wake in front of the obstacle is more
effective in the Gaussian case than for the $\delta$-impurity and is
very well described by the asymptotic form \eqref{eq:wake-long-dist}
as shown in the lower right panel of Fig.~\ref{fig5}.

\subsection{Drag force}\label{subdrag}

In order to discuss the precise influence of the finite lifetime of
the polaritons on the possible superfluidity of the flow, it is
interesting to compute the drag force $F_d$ experienced by the obstacle. 
$F_d$ is defined as \cite{pavloff02}
\begin{equation}\label{eq:drag}
F_d = \int_{\mathbb{R}} {\rm d}x \, |\psi(x,t)|^2 \,
\partial_x U_{\rm ext}(x,t).
\end{equation}
A natural way to compute $F_d$ is to insert the perturbative
expression \eqref{eq:generic} for $\delta\rho$ in
Eq.~\eqref{eq:drag} (see, e.g., Ref.~\onlinecite{Ast04}). Another convenient way is to
use the stress tensor $T(x,t)$ in a manner similar to what has been done in
Ref.~\onlinecite{pavloff02}. The stress tensor is defined as
\begin{equation}\label{eq:stress1}
T(x,t)=-\,\mbox{Im}(\psi^*\psi_t)
+\tfrac{1}{2}|\psi_{xx}|^2 - \tfrac{1}{2}\rho^2 -\rho\, U_{\rm ext}.
\end{equation}
It verifies the ``non-conservation'' equation
\begin{equation}\label{eq:stress2}
J_t + T_x + \rho(U_{\rm ext})_x
=2\eta \, (1-\rho) \, J,
\end{equation}
where in dimensionless units the momentum current-density $J$ 
coincides with the
particle current-density: $J(x,t)\equiv j(x,t)$. In presence of
damping, in stationary regime, integrating this expression over
position, one gets
\begin{equation}\label{eq:stress3}
F_d = 2\eta \int_{\mathbb{R}} {\rm d}x \, (1-\rho) \, J.
\end{equation}
Within the perturbative approach one can show that $J(X=x+Mt)=-M\,
\delta\rho(X)-2\eta\int_{-\infty}^X{\rm d}y\,\delta\rho(y)$, and using
the result \eqref{eq:cons_mass1} this yields, for an obstacle of type
\eqref{eq:movpot},
\begin{equation}\label{eq:stress4}
\begin{split}
F_d & = 2\eta \, M \int_{\mathbb{R}} {\rm d}x \left[ \delta\rho(x) \right]^2 \\
    & = 2\eta \, M
\int_{\mathbb{R}} \frac{{\rm d}q}{2\pi} \, |\chi(q,-Mq)|^2 \,
|\hat{f}_{\rm ext}(q)|^2.
\end{split}
\end{equation}
We emphasize that \eqref{eq:drag} is generally valid, that
\eqref{eq:stress3} is only valid for a stationary regime in presence
of damping for an obstacle moving at constant velocity,
and that \eqref{eq:stress4} is the perturbative evaluation of
\eqref{eq:stress3}.

\begin{figure}
\includegraphics*[width=0.99\linewidth]{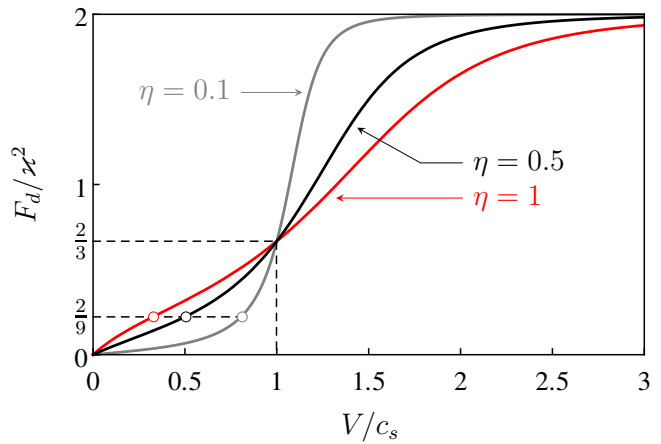}
\caption{(Color online) $F_d/\varkappa^2$ as a function of $M=V/c_s$
  for different values of the dimensionless damping parameter
  $\eta$. The curves are drawn for the $\delta$-impurity potential
  \eqref{eq:7-1}: $U_{\rm ext}(X)=\varkappa\,\delta(X)$.}
\label{fig3}
\end{figure}

For concreteness we now give the explicit
expression of the  perturbative drag \eqref{eq:stress4} in
the case where the potential is a Dirac peak of
the form \eqref{eq:7-1}. One gets
\begin{equation}\label{eq:drag2}
F_d = - \frac{\varkappa^2}{2}
\sum_{\ell\in\{1,2,3\}} \mbox{sgn}(\mbox{Im}\,q_\ell) \, q_\ell \,
\mbox{Res}(q_\ell).
\end{equation}
Substitution of the explicit expressions for the poles yields
\begin{equation}\label{drSub}
F_d=\frac{\varkappa^2\eta\,M(1-M^2)^{-3/2}}{\cos\frac{\theta}{3}
\,(\cos\frac{\theta}{3}+
\frac{1}{\sqrt{3}}\sin\frac{\theta}{3})\,(\cos\frac{\theta}{3}+
\sqrt{3}\sin\frac{\theta}{3})}
\end{equation}
for $M<M_{\rm crit}$ and
\begin{equation}\label{drSuper}
F_d=\frac{8\,\varkappa^2\eta \, M}{F(E^2+9F^2)}
\end{equation}
for $M>M_{\rm crit}$ [in the above expressions $\theta$, $E$ and $F$
are given by Eqs.~\eqref{eq:ap1} and \eqref{eq:ap3old}]. The
behavior of $F_d$ as a function of $M$ is displayed in
Fig.~\ref{fig3} for several values of $\eta$. For each $\eta$ the
critical velocity $M_{\rm crit}$ is reached exactly when the drag is
$F_d=2\varkappa^2/9$. The corresponding points are shown as
white dots in the figure. One can also show that for all $\eta$ one
has $F_d=2\varkappa^2/3$ when $M=1$.

From formulas \eqref{eq:drag2}, (\ref{drSub}) and (\ref{drSuper}) one finds
\begin{equation}\label{eq:apdrag2}
F_d \simeq \varkappa^2 \times
\left\{
\begin{array}{cll}
\eta M & \mbox{when} & M \to 0, \\
     2 & \mbox{when} & M \to \infty,
\end{array}
\right.
\end{equation}
in agreement with the main features of Fig.~\ref{fig3}. It is
interesting to notice that the drag force is proportional to $\eta M$
when $M\to 0$ (a similar behavior has already been observed in
Refs.~\onlinecite{Wou10,Ber12}). This means that at low velocity the
obstacle experiences a force which can be identified to a viscous drag
of Stokes type. When $M$ increases and reaches the value
$M=M_{\rm crit}$, a wake begins to be emitted ahead of the
obstacle. It consists of (damped) Cherenkov radiations and one could
say, pursuing the analogy with fluid mechanics, that this marks the
onset of wave resistance. One can push the analogy one step further
and compare the present results with the ones obtained in experimental
studies of the drag force exerted on objects moving at the surface of
several viscous fluids. In such experiments it is typically observed,
as in Fig.~\ref{fig3}, that the transition to the wave drag is
continuous \cite{Bur01}, but also that $F_d$ considered as a function
of $V$ has a quasi-discontinuous behavior for decreasing viscosity
\cite{Bro01}. An exactly discontinuous behavior is typical for the
perturbative drag in superfluids \cite{pavloff02} and is also expected
on the basis of Rapha\"el--de Gennes theory of wave resistance in the
context of capillary-gravity waves at the surface of inviscid fluids
\cite{Rap96}. This discontinuity disappears for finite viscosity
\cite{Mer11}. Moreover, it is interesting to remark that from
Fig.~\ref{fig3} one might erroneously guess (as is sometimes done in
the analysis of fluid mechanics experiments) that the relevant
critical velocity for the onset of wave drag does not depend on
viscosity (i.e., on $\eta$ in our case) and that at finite viscosity
the behavior of $F_d(M)$ is just smoothed around the inviscid value
[$2\varkappa^2\Theta(M-1)$ in our case]. From our analytical analysis
we know that in reality the wave drag sets in at $M_{\rm crit}$ [which is
not equal to the inviscid value $M_{\rm crit}(\eta=0)=1$] and that it
is not possible, when $M\simeq M_{\mathrm{crit}}$ or $1$, to disentangle in the expression of $F_d$ a
viscous component from a wave resistance. This is clear from
Fig.~\ref{fig3} where the onset of wave drag is shown by thick white
dots: at these points $F_d$ remains a smooth function of $M$.

In Fig.~\ref{fig3} all curves merge at $M=1$, and it is intriguing to
remark that the drag for a fixed velocity $M$ larger than unity
decreases for increased damping. This counter-intuitive effect has
already been observed in a study of the motion of nitrogen drops
floating at the surface of a liquid bath \cite{Mer11}. It is explained
by the fact that viscous effects reduce the range of the wake and
accordingly diminish the wave resistance which is the dominant source
of drag when $M>1$ \cite{rem2}.

\begin{figure}
\includegraphics*[width=0.99\linewidth]{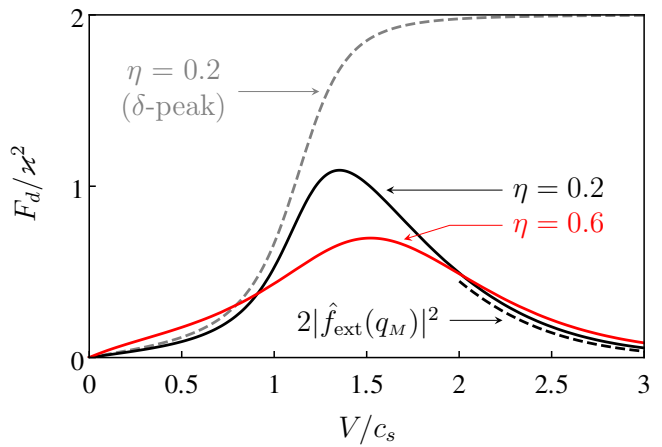}
\caption{(Color online) $F_d/\varkappa^2$ as a function of $M=V/c_s$ for different
  values of the dimensionless damping parameter $\eta$. The solid
  curves are drawn for a Gaussian-impurity potential of width
  $\sigma=0.5$. The black dashed line is the corresponding asymptotic
  result \eqref{eq:drag-large-V}. The gray dashed line is the result
  for a $\delta$-impurity potential, shown for comparison.}
\label{fig4}
\end{figure}

At large velocity all curves in Fig.~\ref{fig3} tend to the same
constant value, which is the result for the drag force in absence of
damping. The fact that the large velocity drag does not depend on $M$
is an artifact of the $\delta$-impurity potential, as demonstrated by
the results obtained in the more standard case where the obstacle is
described by a Gaussian potential of the form \eqref{eq:fw}. In this
case formulas \eqref{eq:drag} or \eqref{eq:stress4} lead to the
expression
\begin{equation}\label{eq:gaussian-drag}
\begin{split}
F_d = - \frac{\varkappa^2}{2}
\sum_{\ell\in\{1,2,3\}} & q_\ell \, \mbox{Res}(q_\ell) \, {\rm e}^{-\sigma^2 q_\ell^2 / 2} \\
& \times \left[\mbox{sgn}(\mbox{Im}\,q_\ell) +
\mbox{erf}\left(\frac{{\rm i} \sigma q_\ell}{\sqrt{2}}\right)
\right].
\end{split}
\end{equation}
The corresponding curves are shown in Fig.~\ref{fig4}. The
counter-intuitive $\eta$-dependence already observed in the case of a
$\delta$-impurity potential is here even more striking: the maximum
drag is larger at small $\eta$ (compare the curves obtained for
$\eta=0.2$ and $\eta=0.6$).

In order to better understand the large-velocity behavior of the
perturbative estimate of the drag force we now derive an explicit
asymptotic expansion valid for any potential of the form
\eqref{eq:movpot} moving at large velocity. From expressions
\eqref{eq:generic} and \eqref{eq:drag} one gets
\begin{equation}\label{eq:b1}
\begin{split}
F_d & = - {\rm i} \int_{\mathbb{R}}
\frac{{\rm d}q}{2\pi} \, q \, \chi(q,-Mq) \, |\hat{f}_{\rm ext}(q)|^2 \\
& = - {\rm i} \int_{\mathbb{R}^2}
{\rm d}x \, \frac{{\rm d}q}{2\pi} \, q \, \chi(q,-Mq) \,
f_{\rm ext} \! \circ \! f_{\rm ext}(x) \, {\rm e}^{-{\rm i} q x}.
\end{split}
\end{equation}
In Eq.~\eqref{eq:b1} $f_{\rm ext}\circ f_{\rm ext}$ is the
convolution of $f_{\rm ext}$ with itself. The in\-te\-gral over $q$ in
this formula can be evaluated by the method of residues. For positive
(negative) $x$ the contour has to be closed from below (above).
Considering that when $M>M_{\rm crit}$ the poles $q_1$ and $q_2$ which
lie in the lower half of the complex $q$-plane verify $q_2 = -q_1^*$ and
$\mbox{Res}(q_2) = - [\mbox{Res}(q_1)]^*$, one gets
\begin{equation}\label{eq:b2}
\begin{split}
F_d & = - 2 \, \mbox{Re}\left[
q_1 \, \mbox{Res}(q_1)
\int_{0}^\infty \! \! {\rm d}x \, f_{\rm ext} \! \circ \! f_{\rm ext}(x) \,
{\rm e}^{-{\rm i} q_1 x} \right] \\
& + q_3 \, \mbox{Res}(q_3) \int_{-\infty}^0 \! \! {\rm d}x \,
f_{\rm ext} \! \circ \! f_{\rm ext}(x) \, {\rm e}^{-{\rm i} q_3 x}.
\end{split}
\end{equation}
At large velocity one obtains, from Eqs.~\eqref{eq:res} and \eqref{eq:ap7},
\begin{equation}\label{eq:b3}
\begin{split}
& q_1\,\mbox{Res}(q_1) = - 2 +
{\cal O}\left(\frac{\eta M}{(M^2-1)^{3/2}}\right), \\
& q_3\,\mbox{Res}(q_3) = {\cal O}\left(\frac{\eta M}{(M^2-1)^{3/2}}\right).
\end{split}
\end{equation}
From this, and using the
fact that $f_{\rm ext}\! \circ\! f_{\rm ext}$ is an even function of $x$,
one can cast the lea\-ding-or\-der contribution to $F_d$ in Eq.~\eqref{eq:b2}
under the form
\begin{equation}\label{eq:b5}
\begin{split}
F_d & = 2
\int_{\mathbb{R}} {\rm d}x \, {\rm e}^{-{\rm i} {\rm Re}(q_1) x} \, {\rm e}^{{\rm Im} (q_1) |x|}
\, f_{\rm ext} \! \circ \! f_{\rm ext}(x) \\
& = 2 \int_{\mathbb{R}} \frac{{\rm d}q}{2\pi} \,
\frac{-2 \, {\rm Im} \, q_1}{[{\rm Re}(q_1) - q]^2 + {\rm Im}^2q_1} \,
|\hat{f}_{\rm ext}(q)|^2.
\end{split}
\end{equation}
The last expression in Eq.~\eqref{eq:b5} is obtained using
Par\-se\-val\---Plancherel theorem. At large velocity the imaginary
part of $q_1$ is of order $\eta M(M^2-1)^{-1}$, whereas its real part is ${\rm
  Re}\,q_1 \simeq q_{\sss M}\equiv2(M^2-1)^{1/2}$ [cf.~Eq.~\eqref{eq:ap7}]: the
Lorentzian in \eqref{eq:b5} is thus a good approximation of the Dirac
distribution $\delta(q-q_{\sss M})$. This directly yields the following large
velocity result:
\begin{equation}\label{eq:drag-large-V}
F_d = 2 \, |\hat{f}_{\rm ext}(q_{\sss M})|^2
\left[ 1 + {\cal O}\left(\frac{\eta M}{(M^2-1)^{3/2}}\right)\right].
\end{equation}
This means that the typical drag depends on velocity (through $q_{\sss
  M}$) and tends to zero at large velocity \cite{rem3} contrarily to
what occurs for the $\delta$-impurity obstacle. It is interesting to
notice that the result \eqref{eq:drag-large-V} does not depend on
$\eta$ at leading order, i.e., that the large-velocity drag
corresponds to pure wave-resistance. Besides, as already remarked in
Sec.~\ref{flowpattern}, the obstacle can always be treated as a
perturbation at large velocity and the associated drag force decreases
(the large velocity limit was accordingly denoted as ``quasi-ideal''
in Ref.~\onlinecite{Alb08}).

\section{Nonlinear theory for a narrow obstacle}\label{nonlinear}

In this section we present results valid for strong obstacle
potentials, in regimes where the perturbative approach of the previous
section typically fails. In the limit of small damping ($\eta\ll 1$)
one can expect that other approximations are valid. For example, in
the case of an obstacle represented by a strong $\delta$-potential,
one can assume that the condensate is strongly disturbed at the
location of the obstacle, so that the difference $1-\rho(0)$ is not small; however, the derivative of the
distribution $\rho(x)$ downstream the obstacle (for $x>0$) is
controlled by $\eta$ and can be considered as small in the case of
small damping. Hence we can develop for this region a so-called {\it
  hydraulic} approximation by neglecting higher order dispersive effects in
our equations (see, e.g., Ref.~\onlinecite{Kam12}). On the other hand,
upstream the obstacle (in the region $x<0$) a supercritical flow
generates a stationary oscillatory structure whose oscillation's
amplitudes are not small, contrarily to what was assumed in the previous
section.  However, in the case of small $\eta$ this oscillatory
structure can be represented as a slowly modulated nonlinear wave and,
hence, the Whitham modulation theory can be applied to its
description. In this section we shall use these two approximate
methods (hydraulic approximation and Whitham averaging technique) and
compare their results with the exact numerical solution of the problem.

In all this section we restrict ourselves to the {\it stationary
  version} of Eq.~\eqref{eq:4-1} in presence of a
$\delta$-impurity. We find it more convenient to work in a reference
frame where the obstacle is at rest while the condensate moves from
left to right with an asymptotic velocity and density respectively
equal to $M$ and $1$ at both infinities.  The equation to be solved is
the following:
\begin{equation}\label{eq:nonlinear1}
\big(\tfrac{M^2}{2}+1\big) \psi=-\tfrac{1}{2}\psi_{xx}+\big(\varkappa 
\,\delta(x)+ \rho\big)\psi+{\rm i}\eta\big(1-\rho\big)\psi.
\end{equation}
Contrarily to the case of a weak obstacle, where one can show that a
stationary solution always exists within perturbation theory (see
Sec.~\ref{3a}), it is not {\it a priori} evident that
Eq.~\eqref{eq:nonlinear1} admits a solution. Hence, the assumption of
existence of a stationary nonlinear regime has to be validated by
exhibiting the corresponding solution and demonstration of its
stability.  If such a solution cannot be found, this means that only
time-dependent flows exist for the chosen values of $\eta$,
$\varkappa$ and $M$, which are the three parameters characterizing
the flow.

By means of the substitution
\begin{equation}\label{k-2}
    \psi(x)=\sqrt{\rho(x)}\exp\left[ {\rm i} \int^x {\rm d}x' \, u(x')\right],
\end{equation}
the Gross--Pitaevskii equation \eqref{eq:nonlinear1} can be cast---outside 
the range of action of the obstacle potential---into a
hy\-dro\-dyna\-mic form for the rescaled density $\rho(x)$ and flow
velocity $u(x)$:
\begin{equation}\label{k-3}
    \begin{split}
    & (\rho \, u)_{x} = 2\eta \, \rho \, (1-\rho), \\
    & \frac{u^2}{2} + \rho + \frac{\rho_{x}^2}{8\rho^2}
    -\frac{\rho_{xx}}{4\rho} = \frac{M^2}2+1.
    \end{split}
\end{equation}
We shall use these hydrodynamic notations in this section.

\subsection{Hydraulic approximation in the downstream
  region of a supersonic flow}\label{hydraulic}

In the hydraulic approximation the derivatives are supposed to be
small; hence we can neglect the two last terms in the left-hand side
of the second of Eqs.~(\ref{k-3}) to get $u^2/2+\rho=M^2/2+1$. Then
$u(x)$ can be expressed in terms of $\rho(x)$ and substituted into the
first of Eqs.~(\ref{k-3}) to give
\begin{equation}\label{d3}
\left[\rho\sqrt{M^2+2(1-\rho)}\right]_x=2\eta\,\rho\,(1-\rho).
\end{equation}
The solution of this equation, with the boundary condition
\begin{equation}\label{d4}
\rho(0)\equiv\bar{\rho},
\end{equation}
can be easily expressed in terms of elementary functions:
\begin{equation}\label{d5}
\begin{split}
x & = \frac{1}{2\eta}
\Bigg \{ \left( M - \frac{1}{M} \right) \\
& \times \ln \frac{ (1-\bar{\rho}) \left[ M^2+1 - \rho + M\sqrt{M^2+2(1-\rho)} \right] }
{ (1-\rho) \left[ M^2+1 - \bar{\rho} + M\sqrt{M^2+2(1-\bar{\rho})} \right] } \\
& - \sqrt{M^2+2} \\
& \times
\ln \frac{\bar{\rho}
\left[ M^2+2 - \rho + \sqrt{(M^2+2)(M^2+2(1-\rho))} \right]}
{\rho
\left[ M^2+2 - \bar{\rho} + \sqrt{(M^2+2)(M^2+2(1-\bar{\rho}))} \right]}
\Bigg \}.
\end{split}
\end{equation}
This formula implicitly defines the dependence of the density $\rho$
on $x$.

In the supersonic case, in the far downstream region, one has
$1-\rho(x)\ll 1$ and one can linearize Eq.~(\ref{d3}) with respect to
$\delta\rho=\rho-1$. This yields\cite{Kam12}
\begin{equation}\label{d6}
|\delta\rho(x)|\propto \exp\left(-\frac{2\eta M}{M^2-1}\,x\right).
\end{equation}
The perturbation theory used in the previous section predicts the same
behavior when $\eta M(M^2-1)^{-3/2}\ll 1$ [$\delta\rho$ is found to be
proportional to $\exp({\rm i} q_3 x)$, where $q_3$ is given by
\eqref{eq:ap7}]. However, the range of validity of Eq.~(\ref{d6}) is
different: the condition of smallness of the derivative yields the
following condition of applicability of the hydraulic approximation:
\begin{equation}\label{d7}
\frac{\eta M}{M^2-1}\ll 1.
\end{equation}
As a consequence of these different regimes of validity one can make
the following remark: if $1-\bar{\rho}\ll 1$, the linearization of
Eq.~(\ref{d3}) can be extended down to $x=0$, yielding $\rho(x\geqslant
0)\simeq 1-(1-\bar{\rho})\exp[-2\eta M x/(M^2-1)]$. As we shall
see in the numerical section \ref{numerics}, this approximation has a
larger range of validity than the pure perturbation approach of
Sec.~\ref{perturbation}. This larger range of validity of the
linearized version of (\ref{d5}) is a result of a drawback of the
hydraulic approximation: the value of $\bar{\rho}=\rho(0)$ is not
predicted by this method and has to be specified before comparison
with numerical results. However, we will see in Sec.~\ref{numerics}
that once this is done, Eq.~(\ref{d5}) gives an excellent account of
the downstream wave-pattern with slow gradients in a supersonic
flow\cite{remh}.

\subsection{Whitham approximation in the upstream
  region of a supersonic flow}\label{whitham}

Upstream the obstacle (when $x<0$) supercritical flows typically
generate a dispersive shock wave which is the nonlinear version of the
oscillatory wake observed in Sec.~\ref{perturbation}.  Now the
amplitude of this wave cannot be considered as small, but for small
$\eta$ its parameters are poorly modified over one
wavelength. Therefore we can describe such a flow within Whitham
modulation theory which is a nonlinear adiabatic approach
\cite{whitham74}.

The nonlinear progressive periodic wave solution can be written in the
form (see, e.g., Refs.~\onlinecite{kamch2000,legk})
\begin{equation}\label{k-4}
\begin{split}
    \rho(x,t) & = \tfrac14 (\la_1-\la_2-\la_3+\la_4)^2 +
    (\la_1-\la_2) (\la_3-\la_4) \\
    & \times \mathrm{sn}^2\left(\sqrt{(\la_1-\la_3)(\la_2-\la_4)}\,(x-V_{\varphi}\,t),m\right)
\end{split}
\end{equation}
and
\begin{equation}\label{k-5}
    u(x,t) = V_{\varphi} + \frac{j}{\rho(x,t)},
\end{equation}
where $\mathrm{sn}$ is the sine elliptic Jacobi function,
\begin{equation}\label{k-6}
   V_{\varphi} = \frac12 \sum_{i=1}^4 \la_i, \qquad
   m = \frac{(\la_1-\la_2)(\la_3-\la_4)}{(\la_1-\la_3)(\la_2-\la_4)},
\end{equation}
and
\begin{equation}\label{k-6b}
\begin{split}
    j & = \tfrac18 (-\la_1-\la_2+\la_3+\la_4) \\
    & \times (-\la_1+\la_2-\la_3+\la_4)
    (\la_1-\la_2-\la_3+\la_4).
    \end{split}
\end{equation}
The parameters $\la_1\leqslant\la_2\leqslant\la_3\leqslant\la_4$ are called the
Riemann invariants of the system.  In the case of strictly periodic
solutions they are constant and they determine characteristics of the
wave such as the phase velocity $V_{\varphi}$ [Eq.~(\ref{k-6})], the
current $j$ evaluated in the frame where the wave is standing [Eq.~\eqref{k-6b}], the
amplitude of the oscillations
\begin{equation}\label{k-7}
    a=(\la_1-\la_2)(\la_3-\la_4),
\end{equation}
and their wavelength
\begin{equation}\label{k-8}
L = \frac{2\,\K(m)}{\sqrt{(\la_1-\la_3)(\la_2-\la_4)}},
\end{equation}
$\K(m)$ being the complete elliptic integral of the first kind. In the
modulated dispersive shock wave occurring in the upstream region, the
$\lambda$'s become functions of $x$ which vary weakly over one
wavelength. We consider here the stationary solution and hence these
parameters do not depend on time $t$ and the phase velocity
$V_{\varphi}$ is equal to zero:
\begin{equation}\label{k-9}
V_{\varphi} = \tfrac12 \sum_{i=1}^4 \la_i = 0.
\end{equation}
However, in the upstream region, the $\lambda$'s are functions of
position and their $x$-dependence is determined by the Whitham
equations (see Appendix \ref{AppC})
\begin{equation}\label{eq:14-1}
\frac{{\rm d}\la_i}{{\rm d}x}=
\frac{2}{L}\,\frac{G_1\la_i+G_2}{\prod_{m\neq i}(\la_i-\la_m)},
\quad i\in\{1,2,3,4\},
\end{equation}
where
\begin{equation}\label{14-2}
\begin{split}
&G_1
=-\eta\int_{\nu_1}^{\nu_2}{\rm d}\nu\,\frac{\nu(1-\nu)}{\sqrt{\mathcal{R}(\nu)}},\\
&G_2=
-\frac{\eta \sqrt{\nu_1\nu_2\nu_3}}{2} 
\int_{\nu_1}^{\nu_2}{\rm d}\nu\,\frac{1-\nu}{\sqrt{\mathcal{R}(\nu)}},
\end{split}
\end{equation}
$\mathcal{R}(\nu)$ and $\nu_1$, $\nu_2$, $\nu_3$ being defined by
Eqs.~(\ref{A-14}) and (\ref{A-15}).  According to Eq.~(\ref{k-9}), the
system \eqref{eq:14-1} admits the first integral $\sum_{i=1}^4\lambda_i=0$.
We shall now show that it admits another integral and can thus be
reduced to a set of two (coupled) differential equations. To this end,
we shall use the Jacobi identities
\begin{equation}\label{14-3}
\begin{split}
&\sum_{i=1}^4\frac{\la_i^k}{\prod_{m\neq i}(\la_i-\la_m)}=0
\quad \text{for} \quad 0 \leqslant k \leqslant 2, \\
\mbox{and} \quad &\sum_{i=1}^4\frac{\la_i^3}{\prod_{m\neq i}(\la_i-\la_m)}=1,
\end{split}
\end{equation}
to obtain $\frac{{\rm d}s_1}{{\rm d}x}=0=\frac{{\rm d}s_2}{{\rm d}x}$ and
\begin{equation}\label{14-4}
\frac{{\rm d}s_3}{{\rm d}x}=\frac{2G_1}L, \qquad
\frac{{\rm d}s_4}{{\rm d}x}=-\frac{2G_2}L,
\end{equation}
where the $s$'s are symmetric functions of the $\la$'s:
\begin{equation}\label{14-4b}
\begin{array}{ll}
\vspace{2mm}
\displaystyle{s_1=\sum_i\lambda_i,} &
\displaystyle{s_2=\sum_{i< j} \lambda_i \lambda_j,} \\
\displaystyle{s_3=\sum_{i< j< k}\lambda_i \lambda_j \lambda_k,\quad} &
\displaystyle{s_4=\lambda_1 \lambda_2 \lambda_3 \lambda_4.}
\end{array}
\end{equation}
Here $s_1$ and $s_2$ are the integrals of our system.  The value of $s_1$
is already known from Eq.~(\ref{k-9}): $s_{1}=0$. In order to determine the value
of $s_2$ we calculate the asymptotic values of the Riemann invariants
at $x\to-\infty$, where the flow is stationary with $\rho=\rho_0=1$ and
$u=u_0=M>0$. The amplitude of the oscillations vanishes here; hence
we find from (\ref{k-7}) that $\la_1=\la_2$ (another possible choice
is $\la_3=\la_4$; it corresponds to a flow with $M<0$). Then, from 
Eq.~(\ref{A-15}) we have the equation
\begin{equation}\label{14-5}
\begin{split}
\underset{x\to-\infty}{\lim}\rho(x)&=\rho_0=1 \\
&=\nu_1=\nu_2=\tfrac14(\la_3-\la_4)^2,
\end{split}
\end{equation} 
as well as the expression for the current density
\begin{equation}\label{14-5a}
\begin{split}
\underset{x\to-\infty}{\lim}j(x) &= \rho_0u_0 = M \\
              &=\tfrac18(\la_3-\la_4)^2(-2\la_1+\la_3+\la_4),
\end{split}
\end{equation}
from which we get another equation:
\begin{equation}\label{14-6}
M=\tfrac12(-2\la_1+\la_3+\la_4).
\end{equation}
With account of Eq.~(\ref{k-9}) (that is $2\la_1+\la_3+\la_4=0$) we
find, at $x\to-\infty$,
\begin{equation}\label{15-2}
\la_1 = \la_2 = -\frac{M}2, \quad \la_3=
\frac{M}2-1, \quad \la_4=\frac{M}2+1.
\end{equation}
Hence,
\begin{equation}\label{15-3}
s_2=\mathrm{C}\textsuperscript{st}=-\frac{M^2}2-1.
\end{equation}
As a result, we can define the functions $\la_i=\la_i(s_3,s_4)$ as being the
roots of the equation
\begin{equation}\label{15-4}
\la^4-\left(\frac{M^2}2+1\right)\la^2-s_3\,\la+s_4=0,
\end{equation}
ordered according to $\la_1\leqslant\la_2\leqslant\la_3\leqslant\la_4$. Substitution of these
functions into (\ref{A-15}) and of the results into (\ref{14-4}) yields
the system of two differential equations for $s_3$ and $s_4$,
\begin{equation}\label{15-5}
 \frac{{\rm d}s_3}{{\rm d}x}=\frac{2G_1(s_3,s_4)}{L(s_3,s_4)},
\quad \frac{{\rm d}s_4}{{\rm d}x}=-\frac{2G_2(s_3,s_4)}{L(s_3,s_4)}.
\end{equation}
We now have to find the initial conditions for this system, that is to
determine the values of $s_3$ and $s_4$ at $x=0$. To this end, we take
into account that Whitham theory implies that the parameters of the wave
weakly change over a distance of about one wavelength. Therefore we can
assume that, to the left of the obstacle and close enough to it, the
wave can be approximated by the cnoidal wave solution \eqref{k-4},
\eqref{k-5} and to the right of the obstacle it is given by a
hydraulic approximation parameterized by the the value $\bar{\rho}$
of the density at the location of the $\delta$-obstacle.

It is known (see, e.g., Ref.~\onlinecite{kamch2000}) that a
non-modulated cnoidal wave solution $\rho(x)$ satisfies the equation
\begin{equation}\label{w1}
\rho_x=2\sqrt{\mathcal{R}(\rho)},
\end{equation}
where the coefficients of the polynomial 
\begin{equation}\label{15-6}
\begin{split}
\mathcal{R}(\nu)& =(\nu-\nu_1)(\nu-\nu_2)(\nu-\nu_3)\\
& =\nu^3+2s_2\nu^2+(s_2^2-4s_4)\nu-s_3^2\; 
\end{split}
\end{equation}
are expressed in terms of the symmetric functions $s_2$, $s_3$ and $s_4$.
Then the solution \eqref{k-4}, \eqref{k-5} (with $V_{\varphi}=0$)
can be expressed in terms of the zeros $\nu_1$, $\nu_2$, $\nu_3$ 
of this polynomial as follows
\begin{equation}\label{w3}
\begin{split}
\rho(x)&=\nu_1+(\nu_2-\nu_1)\,\mathrm{sn}^2\left(\sqrt{\nu_3-\nu_1}\,x,m\right),\\
u(x)&=\frac{s_3}{\rho(x)},
\end{split}
\end{equation}
where
\begin{equation}\label{w4}
m=\frac{\nu_2-\nu_1}{\nu_3-\nu_1}\quad\textrm{and}\quad L=
\frac{2\,\mathrm{K}(m)}{\sqrt{\nu_3-\nu_1}}.
\end{equation}
In the stationary modulated situation we consider that $\nu_1$, $\nu_2$,
$\nu_3$, $s_3$, $m$ and $L$ do not depend on time in Eqs.~\eqref{w3}
and \eqref{w4}, but they all depend on $x$.
 
It follows from Eq.~\eqref{eq:cons_mass} that the current of
polaritons is preserved in transition through the $\delta$-potential:
$j(0^-)=j(0^+)$. Then the second of Eqs.~(\ref{w3}) yields (under the
assumption that the hydraulic approximation is valid for $x\geqslant 0$ because
$\eta\ll 1$) the value of $s_3(0)$:
\begin{equation}\label{w5}
s_3(0)=u(0)\rho(0)=\bar{\rho}\sqrt{M^2+2(1-\bar{\rho})}. 
\end{equation}

For calculating the value of $s_4(0)$ we use the matching condition at $x=0$:
\begin{equation}\label{16-3}
\rho_x(0^+)-\rho_x(0^-)=4\varkappa\rho(0).
\end{equation}
Pursuing the use of the downstream hydraulic approximation already
used in Eq.~(\ref{w5}) we write $\rho(0)=\bar{\rho}$ and, from
Eq.~(\ref{d3}),
\begin{equation}\label{add1}
\rho_x(0^+)=
\frac{2\eta\,\bar{\rho}\,(1-\bar{\rho})\sqrt{M^2+2(1-\bar{\rho})}}
{M^2+2-3\bar{\rho}}.
\end{equation}
In the same spirit of a small-$\eta$ approximation
we have from Eq.~(\ref{w1})
$\rho_x(0^-)=-2\sqrt{\mathcal{R}(\bar{\rho})}$, so that Eq.~(\ref{16-3}) reads
\begin{equation}\label{add2}
\mathcal{R}(\bar{\rho})=\left[2\varkappa\bar{\rho}-
\frac{\eta\,\bar{\rho}\,(1-\bar{\rho})\sqrt{M^2+2(1-\bar{\rho})}}
{M^2+2-3\bar{\rho}}\right]^2.
\end{equation}
This yields
\begin{equation}\label{16-5}
\begin{split}
& (\nu_1\nu_2+\nu_2\nu_3+\nu_3\nu_1)_{x=0} = s_2^2-4s_4(0) \\
& = \frac{[4\varkappa\bar\rho-\rho_x(0^+)]^2}{4\bar\rho}+
(2M^2+4-3\bar{\rho})\bar{\rho},
\end{split}
\end{equation}
and then
\begin{equation}\label{add3}
\begin{split}
s_4(0)&=
\frac14\left[\frac{(M^2+2)^2}{4}-2(M^2+2)\bar{\rho}+3\bar{\rho}^2\right]\\
&-\bar{\rho}\left[\varkappa-
\frac{\eta\,(1-\bar{\rho})\sqrt{M^2+2(1-\bar{\rho})}}{2(M^2+2-3\bar{\rho})}
\right]^2.
\end{split}
\end{equation}
We note here that for small values of $\eta$, the analytical
expression (\ref{add3}) can be
simplified by replacing Eq.~(\ref{add1}) by the simple approximation
$\rho_x(0^+)=0$.  This amounts to also replace $\eta$ by 0 in the
expressions (\ref{add2}) and (\ref{add3}).
This simple scheme is accurate when
$\eta\lesssim 0.5$.

Equations (\ref{w5}) and (\ref{add3}) give the initial
conditions for the system
\begin{equation}\label{17-1}
\begin{split}
\frac{{\rm d}s_3}{{\rm d}x}&=-
\frac{2\,\eta}{L}
\int_{\nu_1(s_3,s_4)}^{\nu_2(s_3,s_4)}{\rm d}\nu\,
\frac{\nu(1-\nu)}{\sqrt{\mathcal{R}(\nu)}}, \\
\frac{{\rm d}s_4}{{\rm d}x}&=
\frac{\eta \, s_3}{L}
\int_{\nu_1(s_3,s_4)}^{\nu_2(s_3,s_4)}{\rm d}\nu\,
\frac{1-\nu}{\sqrt{\mathcal{R}(\nu)}},
\end{split}
\end{equation}
where $\nu_i(s_3,s_4)$ ($i=1,2,3$) are determined as being the roots
of the equation $\mathcal{R}(\nu,s_3,s_4)=0$, where
\begin{equation}\label{17-2}
\begin{split}
\mathcal{R}(\nu,s_3,s_4) \equiv
\nu^3 &-(M^2+2)\nu^2\\
& +\bigg[\left(\frac{M^2}{2}+1\right)^2-4s_4\bigg]\nu-s_3^{2}.
\end{split}
\end{equation}
In Eqs.~(\ref{17-1}) $L$ is also expressed in terms of the $\nu$'s
[see Eqs.~(\ref{w4})].

In the present application of Whitham modulation theory it is
important to notice that for fixed values of $\varkappa$, $\eta$ and
$M$, the solution of Whitham equations depends on a single parameter
$\bar{\rho}$ which is also a function of the same set of physical
parameters ($\varkappa,\,\eta,\,M$) prescribed by the external
potential and the boundary conditions of the Gross--Pitaevskii
equation. Hence, the parameter $\bar{\rho}$ can be found from the
condition that the solution of Whitham equations satisfies the
correct boundary condition at $x\to-\infty$, namely that the
envelopes of the density
oscillations tend to the asymptotic value of the density:
\begin{equation}\label{add4}
  \nu_1(x),\,\nu_2(x)\to 1 \quad\text{as}\quad x\to-\infty.
\end{equation}
Some values of $\bar{\rho}$ calculated in this way are listed in the second
row of Table \ref{table} for $\eta=0.05$, $M=3$ and several values of
$\varkappa$. We compare them with the values of $\bar{\rho}$ obtained by
exact numerical solution of Eq.~\eqref{eq:nonlinear1}. 
As we see, the agreement is very good.

\begin{figure}
\includegraphics*[width=0.99\linewidth]{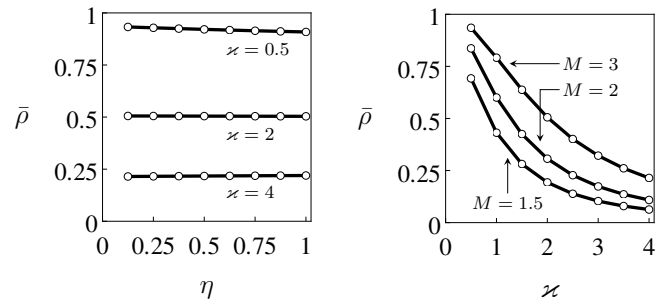}
\caption{$\bar\rho=\rho(0)$ as a function of $\eta$ for
  $M=3$ and different values of $\varkappa$ (left panel) and as a function
  of $\varkappa$ for $\eta=0.05$ and several values of $M$ (right panel).}
\label{fig-RhoBar}
\end{figure}

The dependence of $\bar{\rho}$ on $\eta$ is displayed in
Fig.~\ref{fig-RhoBar} (left panel) for several values of $\varkappa$. This
plot suggests that $\bar{\rho}=\rho(0)$ does not tend to unity in the limit
$\eta\to0$. This means that, in this limit, the flow pattern does not
reduce to the exact solution found in Ref.~\onlinecite{lp-2001} in
the case $\eta\equiv0$, since for this solution $\rho(0)=1$. Rather,
however small is $\eta$, $1-\bar{\rho}$ remains finite, the wave
structure occupies a portion of space proportional to $\eta^{-1}$ and
decays towards an undisturbed flow ($\rho\equiv 1$) at $|x|\gg \eta^{-1}$.
The dependence of $\bar{\rho}$ on $\varkappa$ for several values of $M$
is shown in the right panel of Fig.~\ref{fig-RhoBar}.

A striking feature of the plot in the left panel of Fig.~\ref{fig-RhoBar} is the
extremely weak $\eta$-dependence of $\bar{\rho}$. This important property
of the theory can be explained by the simple fact that the space
coordinate $x$ and the parameter $\eta$ enter into both the hydraulic approximation and Whitham
equations only through the combination $\eta\,x$ [see Eqs.~\eqref{d5} and \eqref{17-1}]. As a result,
$\eta$ can be rescaled out of the exact relation \eqref{eq:cons_mass1} after averaging over fast oscillations
in the dispersive shock region $x<0$, so that we arrive to an equation
which depends on $\eta$ only through the small value of $\rho_x(0^+)$
[see Eqs.~\eqref{add1} and \eqref{add3}]. If we neglect this term, then the resulting
equation yields $\bar{\rho}$ as a function of $M$ and $\varkappa$ only.

\begin{table}
\begin{center}
\begin{tabular}{ c | c  c  c  c  c }
  \hline
  $\varkappa$ & 0.5 & 1.0 & 1.5 & 2.0 & 2.5  \\
  \hline
  $\bar{\rho}$ (Whitham) & 0.9370 & 0.7932 & 0.6384 & 0.5056 & 0.4011 \\
  \hline
  $\bar{\rho}$ (numerics) & 0.9352 & 0.7916 & 0.6377 & 0.5055 & 0.4013 \\
  \hline  
\end{tabular}
\end{center}
\caption{Values of $\bar{\rho}$ for different values of $\varkappa$ in 
  the case $M=3$ and $\eta=0.05$. The row $\bar{\rho}$ (Whitham) 
  corresponds to the 
  value of $\bar\rho=\rho(0)$ found by solving Whitham equations 
  (\ref{17-1}) and imposing the condition (\ref{add4}) (see the text). The row
  $\bar{\rho}$ (numerics) corresponds to the value of $\rho(0)$ 
  found \textit{via} a numerical resolution of Eq.~\eqref{eq:nonlinear1}.}
\label{table}
\end{table}

When $\bar{\rho}$ is found, all the parameters of the dispersive shock
wave are determined, the functions $\nu_1(x)$, $\nu_2(x)$, $\nu_3(x)$
can be computed by solving Eq.~(\ref{17-1}), 
and their substitution into Eq.~(\ref{w3})
yields the oscillatory structure upstream the obstacle. The same value
of $\bar{\rho}$ determines the hydraulic solution downstream the
obstacle. Thus, we reach a complete description of the nonlinear
wave generated by a supercritical flow past a
$\delta$-obstacle.

\begin{figure}
\includegraphics*[width=0.99\linewidth]{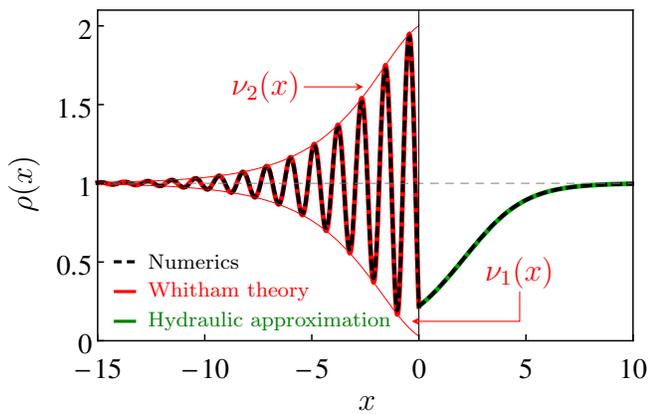}
\caption{(Color online) Comparison of the Whitham theory with the
  numerical solution of Eq.~(\ref{eq:nonlinear1}). The plot is drawn
  in the case $\eta=1$, $M=3$ and $\varkappa=4$. The numerics
  corresponds to the dashed black line. Whitham envelopes are shown by
  thin red solid lines, and the upstream dispersive shock wave
  oscillatory structure obtained by substitution of the solution of
  the Whitham equations (\ref{17-1}) into Eq.~(\ref{w3}) is shown by a
  red solid line (for $x\leqslant 0$). The downstream ($x\geqslant 0$) hydraulic
  approximation is shown by a green solid line.}
\label{fig-whitham}
\end{figure}

The accuracy of the theory is illustrated by Fig.~\ref{fig-whitham}.
As we see, the agreement between the results of the combined Whitham and hydraulic
approaches and the numerical computations is excellent. Note that
Whitham method is perfectly valid in a regime where the perturbative
theory of Sec.~\ref{perturbation} seriously fails ($|\rho(x)-1|$ is
not small in Fig.~\ref{fig-whitham}). For illustrative reasons we
have chosen a relatively large value of $\eta$ ($\eta=1$): we wanted
to work in a regime where the overall modulations of the oscillating
pattern occur over a characteristic length which is not to large
with respect to the wavelength of the oscillations. As we see, even in
this unfavorable case the agreement with the exact numerical results
is very good.

The solution of the system (\ref{17-1}) exists, and the upstream
pattern can be described as a slowly modulated cnoidal wave, as long
as its initial conditions can be found, that is, as long as the
equation $\mathcal{R}(\nu,s_3(0),s_4(0))=0$ has three real roots. If
$\eta$ is strictly zero, then
$\bar{\rho}=1$, and this equation reads
\begin{equation}\label{17-4}
\nu^3+(M^2+2)\nu^2+(1+2M^2+4\varkappa^2)\nu-M^2=0.
\end{equation}
Two of the roots coalesce and go into the complex plane when the
discriminant of Eq.~(\ref{17-4}) vanishes.  This corresponds to a
boundary between possible parameters in the plane $(\varkappa,M)$
determined by the condition
\begin{equation}\label{17-5}
\varkappa_b^2=\frac1{32}\left[M(M^2+8)^{3/2}+M^4-20M^2-8\right].
\end{equation}
The same boundary was already found in a different analytical form in 
Ref.~\onlinecite{lp-2001} for a non-damped system. In our
problem ($\eta\ne 0$, $\bar{\rho}\neq 1$), this boundary is changed
and can be found by numerically determining when the discriminant of
Eq.~(\ref{17-2}) vanishes. However, when $\eta\ne 0$, as we shall see
in Sec.~\ref{numerics}, new stationary solutions appear when $\varkappa$
gets so large that the upstream flow is not described by a
modulated cnoidal wave, making the determination of the domain of
validity of Whitham approach less crucial than when $\eta=0$.

\subsection{Numerical results}\label{numerics}

\begin{figure}
\includegraphics*[width=0.99\linewidth]{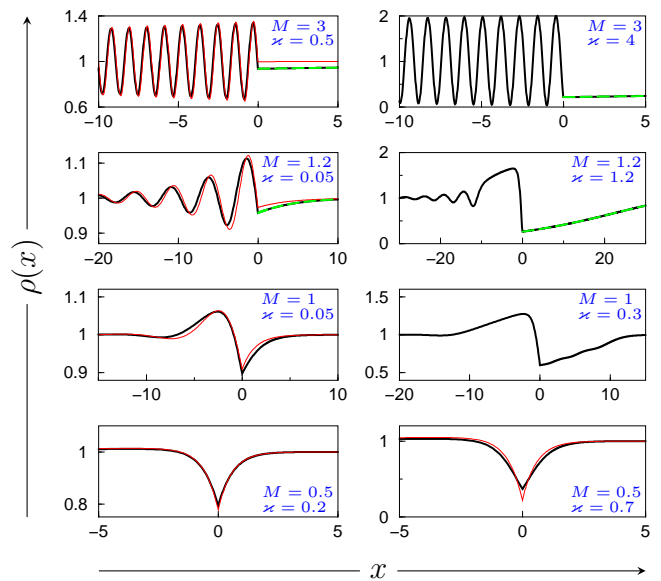}
\caption{(Color online) Different profiles $\rho(x)$ for flows past
  a $\delta$-impurity potential of type \eqref{eq:7-1}. For all the
  profiles the damping parameter is $\eta=0.05$. For the upper row
  $M=3$, then $M=1.2$ for the row below, $M=1$ for the following one
  and finally $M=0.5$ for the lower row. The value of $\varkappa$ is
  indicated in each plot. In each plot the black solid line
  corresponds to the numerical solution of Eq.~\eqref{eq:nonlinear1},
  the (red online) thin line to the perturbative result and the (green
  online) dashed line to the result of the hydraulic approximation
  which is only relevant for $x\geqslant 0$ (see Sec.~\ref{hydraulic}).}
\label{someprofiles}
\end{figure}

In this section we present results of the full numerical solution of
Eq.~\eqref{eq:nonlinear1}.  We used a shooting method,
starting the numerical integration from large and positive $x$ with an initial
behavior given by the prediction of perturbation theory. Typical results
are displayed in Fig.~\ref{someprofiles}.

The upper plots of this figure are drawn for $M=3$ which is a velocity
deep enough in the supersonic regime for Whitham theory of
Sec.~\ref{whitham} to apply over a rather large range of values of
$\varkappa$.  The left plot of the upper row corresponds to
$\varkappa=0.5$. For this value of $\varkappa$, perturbation theory is
valid upstream ($x<0$) but fails for positive $x$, whereas the
hydraulic approximation is quite accurate in this region, as shown by
the dashed line in this plot. For $\varkappa=4$ (right plot of the
upper row of Fig.~\ref{someprofiles}), the density profile shows the
same features, but in this case perturbation theory seriously
fails, whereas the downstream wave pattern being typical
for a damped cnoidal wave is very well described by Whitham
theory (not shown, because undistinguishable from the numerical result).

The two rows below the upper one correspond to $M=1.2$ and $M=1$. They
are interesting because they show that, whereas perturbation theory
fails in absence of damping when $M\simeq 1$, for $\eta\ne 0$ it has
a regime of validity even for velocities $M$ close to unity. This is
illustrated by the good agreement of the perturbative results with
the numerics displayed in the two left plots of the central rows (which
are both drawn in the case $\varkappa=0.05$). It is also
interesting to remark that for $M=1$, no stationary solution exists
when $\eta=0$, whereas here we could find such solutions up to
$\varkappa=0.3$ (see the right plot of the third row). The values
$M=1$ and $\varkappa=0.3$ are close to the boundary marking the end
the existence of stationary solutions when $\eta=0.05$. In this case
we see that the downstream wave pattern shows small scale disturbances
which were recognized in Ref.~\onlinecite{Kam12} as typically occurring
near the end of the stationary regime.

The second upper row of Fig.~\ref{someprofiles} corresponds to $M=1.2$. In this
case, when $\eta=0$, there is no stationary solution for $\varkappa\geqslant
0.0495$ [see Eq.~(\ref{17-5}) or Ref.~\onlinecite{lp-2001}]. As seen on the
figure, when $\eta=0.05$, one can find stationary solutions for much
larger values of $\varkappa$ (up to $\varkappa\simeq 1.2$; see the
corresponding plot). However, the density profile found in this case
is very different from a damped cnoidal wave. It seems to be a
stationary version of a type of time-dependent profiles studied in
Ref.~\onlinecite{Kam12b} for the case $\eta=0$: a plateau develops
just upstream the obstacle which terminates when $x\to -\infty$ by a
dispersive shock wave. Here, when $\eta\ne 0$, the plateau and the
shock wave are damped because the specific form of the modified
Gross--Pitaevskii equation \eqref{eq:4-1} favors relaxation towards
$\rho=1$.

\begin{figure}
\includegraphics*[width=0.99\linewidth]{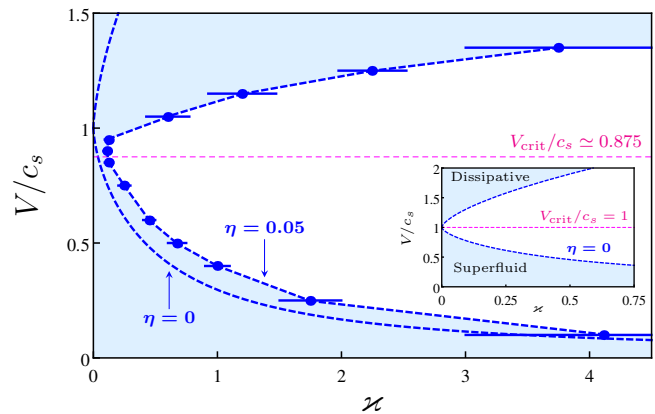}
\caption{(Color online) Different regimes of flow past a
  $\delta$-impurity potential of type \eqref{eq:7-1} in the
  $(\varkappa,M=V/c_{s})$ plane. The main plot corresponds to the
  dissipative system (with $\eta=0.05$). The inset is drawn for
  $\eta=0$. In both plots the shaded regions correspond to stationary
  flows, the white ones to time-dependent and dissipative flows and
  the horizontal dashed line indicated the transition from a localized
  wake to a regime of Cherenkov emission as predicted by perturbation
  theory. In the inset, the subsonic ($M<1$) shaded region is
  superfluid and the supersonic ($M>1$) one is dissipative; the exact
  equation of the boundaries between the different domains is given in
  Ref.~\onlinecite{lp-2001}. In the main plot the dots with error bars
  represent the numerically determined boundary of the stationary
  domain. They are connected by a dashed line to guide the eye. The
  other dashed line in this plot represents the $\eta=0$ result (shown for
  comparison).}
\label{figkm}
\end{figure}

The lower row of Fig.~\ref{someprofiles} displays results
corresponding to a subsonic obstacle ($M=0.5$). For this value of the
velocity, there is no stationary solution in the $\eta=0$ case for
$\varkappa\geqslant 0.59$\cite{lp-2001,Hak97}. As illustrated by the right
plot of this row (drawn for $\varkappa=0.7$) in presence of damping,
solutions exist for slightly larger values of $\varkappa$. However we
find that, when $\eta$ passes from 0 to 0.05, the range of values of
$\varkappa$ allowing for a stationary solution does not increase in
the subsonic case as much as it does in the supersonic region. This is
illustrated by Fig.~\ref{figkm} were we represent the domain of
existence of stationary flows in the $(\varkappa,M)$ plane. This domain
corresponds to the shaded region in the Figure and was numerically
determined in the case $\eta=0.05$.  The large increase of the
stationary domain for supersonic flows in presence of damping is due
to the occurrence, when $\eta\ne 0$, of a new class of profiles with
an upstream plateau, as explained above and illustrated in the right
plot of the second row from the left in Fig.~\ref{someprofiles}
(corresponding to $M=1.2$ and $\varkappa=1.2$). This type of profile
cannot be stationary with the boundary condition
$\rho(x\to\pm\infty)=1$ in a non-dissipative system. Here, the damping
term in Eq.~\eqref{eq:4-1} provides a mechanism allowing the
downstream relaxation from $\rho(0)<1$ to $\rho(x\to\infty)=1$ and the
upstream dispersive shock is stabilized by dissipation.

The inset in Fig.~\ref{figkm} represents the exact domain of
stationary flows for $\eta=0$, as analytically determined in
Ref.~\onlinecite{lp-2001}. One can identify three regimes
depending on the value of the parameters $\varkappa$ and $M=V/c_s$:
(i) subsonic, stationary and superfluid, (ii) dissipative and
time-dependent, (iii) dissipative, stationary and supersonic. As seen
in this inset, regimes (i) and (iii) are always separated by the
time-dependent region (ii). This feature is also valid for a thick
obstacle \cite{legk} and is in contradiction with the (wrong)
prediction of perturbation theory for $\eta=0$. Indeed, in the
non-dissipative case, perturbation theory always fails when $V$ is
close to $c_s$\cite{lp-2001} and in this case the true flow gets
time-dependent. On the contrary, for finite $\eta\ll 1$ we showed in
Sec.~\ref{perturbation} that the perturbative prediction of existence
of a stationary flow pattern {\it for all velocities} is valid until
$\varkappa \sim \eta^{1/3}$. This is corroborated by the numerical
results displayed in Fig.~\ref{figkm} for $\eta=0.05$. In this case
$\eta^{1/3}\simeq 0.3$ whereas the largest value of $\varkappa$ for
which a stationary flow exists for all $M$ is numerically found to be
$\simeq 0.1$. Then, we can go one step further, and since we showed
that the actual small parameter of perturbation theory is $\epsilon=
\varkappa\times\mbox{max}\{1,\eta^{-1/3}\}$ we conjecture that the
neck of the stationary domain in Fig.~\ref{figkm} extends when $\eta$
increases from 0, until $\eta\gtrsim 1$, where the largest value of
$\varkappa$ for which a stationary flow exists for all $M$ should
remain approximatively constant and of order of 1.

\section{Conclusion}\label{conclusion}

In the present work we have analyzed the flow of a one-dimensional
polariton condensate in motion with respect to an obstacle in a
situation of non-resonant pumping. We solved the problem perturbatively
and showed that at this level there exists a smooth crossover from a
viscous flow to a regime where the drag is mainly dominated by wave
resistance. Perturbation theory predicts that this occurs at a velocity
$M_{\rm crit}$ independent of the potential representing the obstacle.
We argued that in the case of a $\delta$-impurity
[represented by a potential of type \eqref{eq:7-1}] the perturbative
approach is valid {\it for all velocities} in the regime
$\varkappa\times \max\{1,\eta^{-1/3}\} \ll 1$, where $\eta$ is the
dimensionless damping parameter defined in Eq.~\eqref{eq:meta}. As
shown in the previous section this implies that stationary profiles
indeed exist for all velocities if $\varkappa\lesssim
\mbox{min}\{1,\eta^{1/3}\}$. In this case there is a continuous
transition from a dissipative drag to a regime dominated by the wave
resistance.

However, from Fig.~\ref{figkm} we are led to refine this
discussion of the transition between a regime where the wake is
localized in vicinity of the obstacle and a regime of (damped)
Cherenkov radiation: we see on the example of the $\delta$-impurity
that for a strong enough potential the two types of flows are separated
by a time-dependent regime, as typically observed in BEC atomic
vapors. In this case one cannot state that the crossover is
smooth.

An important result of our work is the demonstration
that it is difficult to assess on the superfluidity of a polariton
system just by studying the density perturbation past a localized
obstacle. In particular, we showed that the absence of long-range wake
cannot be used as a criterion for the absence of dissipation.

The comparison of our results with the ones of Ref.~\onlinecite{Ber12}
leads to the conclusion that the gross features of the wave pattern
discussed in the present work are quite independent of the technique
used for setting the fluid into motion with respect to the
obstacle. However, we use a specific model [Eq.~\eqref{eq:scal1}] with
non-resonant pumping which is more relevant for the experiment
presented in Ref.~\onlinecite{Amo09a}. In this experiment, a
two-dimensional supersonic cloud of polaritons colliding with an
obstacle was observed to induce a rather well defined wake, with
oscillations having an apparently specified wavelength. The same
feature was observed numerically in Ref.~\onlinecite{Wou10} (see also
the discussion in Ref.~\onlinecite{Car-Cui12}). Our perturbative
results allow to understand this phenomenon in a one-dimensional
setting: the pattern of the upstream oscillatory wake in a
supercritical flow ($V>V_{\rm crit}$) is governed by the complex wave
vectors $q_1$ and $q_2$; see Sec.~\ref{perturbation}. Also in the
nonlinear approach (Whitham theory of Sec.~\ref{whitham}) the
wake keep a simple shape: perturbation theory fails to properly
account for the amplitude of the oscillations, but it still
approximatively describes their wavelength.

Finally, this work naturally calls for developments. One would first
like to precisely determine the domain of time-dependent nonlinear
flows in presence of damping. Secondly, one would like to extend the present work for
taking into account polarization effects, and, thirdly, it is natural to
apply the perturbative approach to higher dimensions. Works in these
directions are in progress.

\begin{acknowledgments}
We thank A. Amo, J. Bloch, M. Rabaud and M. Ri\-chard for fruitful discussions.
  A. M. K. thanks LPTMS (Universit\'e Paris Sud and CNRS), where this
  work was done, for kind hospitality. This work was supported by RTRA
  Triangle de la Physique.
\end{acknowledgments}

\appendix

\section{Poles of the response function $\chi(q,-Mq)$}\label{AppA}

In this appendix we determine---as a function of $M$---the location
in the complex $q$-plane of the poles of the response function
\eqref{eq:Linear-response-function} evaluated at $\omega=-M
q$. Considering the expression of $\chi$ one sees that these poles are
the three zeros of $D(q,-Mq)/q$. We denote them as $q_1$, $q_2$ and
$q_3$. They are solutions of Eq.~(\ref{7-2a}).
This equation has three imaginary solutions when its discriminant
$\Delta = 256 (1-M^2)^3 / 27 - 64 \, \eta^2 M^2$ is positive. The condition
$\Delta>0$ is equivalent to $M<M_{\rm crit}$ where
the expression of $M_{\rm crit}$ is given
in Eq.~\eqref{eq:vcr}. In this case, defining
\begin{equation}\label{eq:ap1}
\theta=\arctan\left(\frac{8\eta M}{\sqrt{\Delta}}\right),
\end{equation}
one finds
\begin{equation}\label{eq:ap2}
\begin{split}
q_1&= 4 {\rm i} \,\sqrt{\frac{1-M^2}{3}}
\sin\left(\frac{\theta}{3}-\frac{\pi}{3}\right),\\
q_2&= - 4 {\rm i} \,\sqrt{\frac{1-M^2}{3}}
\sin\left(\frac{\theta}{3}\right),\\
q_3&= 4 {\rm i} \,\sqrt{\frac{1-M^2}{3}}
\sin\left(\frac{\theta}{3}+\frac{\pi}{3}\right).
\end{split}
\end{equation}
Alternatively one can write $q_1={\rm i}(-A+B)$, $q_2=- 2 {\rm i} B$ and
$q_3={\rm i}(A+B)$ with
\begin{equation}\label{eq:ap2-old}
\begin{bmatrix}A \\ B\end{bmatrix}
= 2 \sqrt{1-M^2}
\begin{bmatrix}\cos(\theta/3) \\ \frac{1}{\sqrt{3}} \sin(\theta/3)
\end{bmatrix}.
\end{equation}

If $\Delta<0$, i.e., if $M>M_{\rm crit}$,
defining 
\begin{equation}\label{eq:ap3a}
D_{(\pm)}=\left(4 \eta M \pm \frac{1}{2} |\Delta|^{1/2}\right)^{1/3},
\end{equation}
one finds
\begin{equation}\label{eq:ap3}
\begin{split}
q_1 & = D_{(+)} \exp(-{\rm i} \pi/6) - D_{(-)} \exp({\rm i} \pi/6), \\
q_2 & = - D_{(+)} \exp({\rm i} \pi/6) + D_{(-)} \exp(-{\rm i} \pi/6), \\
q_3 & = {\rm i} (D_{(+)} + D_{(-)}).
\end{split}
\end{equation}
Alternatively one can
write $q_1=E-{\rm i} F$, $q_2=-E-{\rm i} F$ and $q_3=2 {\rm i} F$ with
\begin{equation}\label{eq:ap3old}
E = \frac{\sqrt{3}}{2} (D_{(+)}-D_{(-)}),\quad
F = \frac{1}{2} (D_{(+)}+D_{(-)}).
\end{equation}
One can verify that $\sum_{\ell=1}^3 q_\ell=0$ for all values of $M$, as
already clear from the form of Eq.~(\ref{7-2a}). A similar relation
holds for the residues of $\chi(q,-Mq)$ whose expressions are given in
\eqref{eq:res}: $\sum_{\ell=1}^3 \mbox{Res}(q_\ell)=0$.

\begin{figure}
\includegraphics*[width=0.99\linewidth]{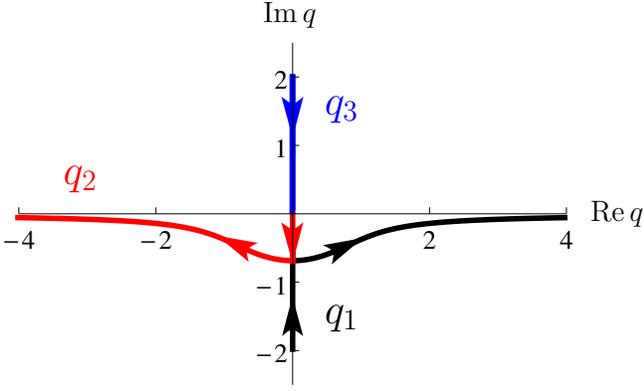}
\caption{(Color online) Position of $q_1$, $q_2$ and $q_3$ in the
  complex $q$-plane. The figure is drawn
  in the case $\eta=0.1$. The arrows indicate the direction
of motion of the
  poles when $M$ increases from $0$ to $\infty$ \cite{rem}.}
\label{fig-app}
\end{figure}

The typical $M$-dependence of the position of the poles in the complex
plane is illustrated in Fig.~\ref{fig-app}. When $M=0$ one has
$\theta=0$, $q_2=0$ and $q_3=-q_1=2 {\rm i}$.  When $M$ is increased from
zero, $q_1$ and $q_2$ get closer on the imaginary axis until they
collide (when $M=M_{\rm crit}$) and then acquire a finite real part.
When $M\to\infty$, $q_3\to {\rm i} \, 0^+$ and
$q_{\left(\begin{smallmatrix}1\\2\end{smallmatrix}\right)}\to
(\pm)\infty-{\rm i} \, 0^+$.

A useful approximation for the expression of the poles is obtained
when $\eta M/|M^2-1|^{3/2}\ll 1$. In this case one obtains, 
when $M<M_{\rm crit}$,
\begin{equation}\label{eq:ap6}
\begin{split}
q_{\left(\begin{smallmatrix}1\\3\end{smallmatrix}\right)}
&\simeq {\rm i} \left[(\mp)2\sqrt{1-M^2}+\frac{\eta M}{1-M^2}\right],\\
q_2&\simeq -2 {\rm i} \, \frac{\eta M}{1-M^2},
\end{split}
\end{equation}
and when $M>M_{\rm crit}$,
\begin{equation}\label{eq:ap7}
\begin{split}
q_{\left(\begin{smallmatrix}1\\2\end{smallmatrix}\right)}
&\simeq (\pm)2\sqrt{M^2-1}-{\rm i} \, \frac{\eta M}{M^2-1},\\
q_3&\simeq 2 {\rm i} \, \frac{\eta M}{M^2-1}.
\end{split}
\end{equation}
The above expressions are valid up to corrections of relative order
$\eta^2M^2/|M^2-1|^3$. It is interesting to notice that expansions
\eqref{eq:ap6} and \eqref{eq:ap7} are equally valid at large velocity
and at small damping. Indeed, as discussed at the end of
Sec.~\ref{subdrag}, at large velocity the effects of damping are
negligible.

From the explicit expressions \eqref{eq:ap2} and \eqref{eq:ap3} of the
$q_\ell$'s it is a simple matter to evaluate the integral \eqref{eq:K}
which permits to compute the function $K(X)$. One gets
\begin{equation}\label{eq:ap9}
\begin{split}
K(X\geqslant 0)&= {\rm i} \, \mbox{Res}(q_3) \,
{\rm e}^{{\rm i} q_3 X}, \\
K(X\leqslant 0)&=- {\rm i} \left[ \mbox{Res}(q_1) \,
{\rm e}^{{\rm i} q_1 X} + \mbox{Res}(q_2) \, {\rm e}^{{\rm i} q_2 X}\right].
\end{split}
\end{equation}
Formulas \eqref{eq:ap9} are valid for all $M$, but the explicit expressions for
the $q_\ell$'s depend on $M$. For instance, when $M<M_{\rm crit}$ the $q_\ell$'s
are all imaginary and $K$ tends rapidly to zero when $|X|\to\infty$.
On the other hand, when $M>M_{\rm crit}$ the exponential decrease of
$K(X)$ gets weaker (because the imaginary part of the $q_\ell$'s is
smaller) and $K(X\leqslant 0)$ oscillates (because $q_1$ and $q_2$ acquire a
real part). The typical density perturbations associated
with $K$ [i.e., for a $\delta$-peak potential of the form
\eqref{eq:7-1}] are sketched in the insets of
Fig.~\ref{fig2}. Note that
the value of the $q_\ell$'s does not depend on $\eta$ when $M=0$, i.e.,
within the theoretical description corresponding to Eq.~\eqref{eq:scal1},
the density perturbation induced by a
motionless obstacle does not depend on the damping.

The expressions \eqref{eq:ap9} are equally valid in absence
of damping, i.e., when $\eta=0$. In this case $M_{\rm crit}=1$,
$q_2=0$ and $q_3=-q_1$ for $M<M_{\rm crit}$ and for $M>M_{\rm crit}$,
$q_3=0$ whereas $q_1$ and $q_2$ are real and opposite
[cf.~Eqs.~\eqref{eq:ap6} and \eqref{eq:ap7}]: for $M>M_{\rm crit}$ and
$\eta=0$ one observes undamped Cherenkov radiations ahead of the
obstacle as discussed in the main text. For $M>M_{\rm crit}$ and
$\eta\ne 0$ these Cherenkov radiations are damped since in this case
$q_1$ and $q_2$ have a nonzero imaginary part.

Finally, we need to evaluate the order of magnitude of the quantity
$\varkappa \, |K(0)|$ at $M=M_{\rm crit}$ since, as argued in the
main text (Sec.~\ref{3b}), this is the small parameter of perturbation
theory for a $\delta$-impurity obstacle.
For $M=M_{\rm crit}$ one gets 
$\theta=\pi/2$, 
$q_1=q_2=-q_3/2=-2{\rm i}\sqrt{(1-M^2_{\rm crit})/3}$
[cf. Eqs.~\eqref{eq:ap2}] and this yields
\begin{equation}\label{eq:kde0}
\varkappa \, K(0) = 
\frac{{\rm i}\, \varkappa\, \sqrt{3}}{\sqrt{1-M^2_{\rm crit}}}
\quad\mbox{when}\quad M=M_{\rm crit}.
\end{equation}
From the expression \eqref{eq:vcr} for $M_{\rm
  crit}$ one sees that $(1-M^2_{\rm crit})^{-1/2}\simeq
\frac{1}{\sqrt{3}} (2/\eta)^{1/3}$ when $\eta\ll 1$ and tends to unity
at large $M$, from which one obtains the estimate \eqref{eq:ap10}.

\section{Derivation of perturbed Whitham equations}\label{AppC}

The general method of derivation of the Whitham equations for
perturbed integrable equations which in their nonperturbed form belong
to the Ablowitz--Kaup--Newell--Segur scheme was developed in Ref.~\onlinecite{kamch2004} 
and it can be formulated as follows. Let the evolution equations of some
field variables $u_k$ have the form
\begin{equation}\label{A-1}
\begin{split}
\frac{\prt u_k}{\prt t}&=
K_{k}\left(u_m,\eps\frac{\prt u_m}{\prt x},\eps^2\frac{\prt^2 u_m}{\prt x^2},
\ldots\right)\\
&+R_{k}\left(u_m,\eps\frac{\prt u_m}{\prt x},\eps^2\frac{\prt^2 u_m}{\prt x^2},
\ldots\right),
\end{split}
\end{equation}
where a small parameter $\eps\ll1$ is introduced which measures the
dispersion effects. It is supposed that a non-perturbed system
\begin{equation}\label{A-2}
\eps\frac{\prt u_k}{\prt t}
=K_k\left(u_m,\eps\frac{\prt u_m}{\prt x},
\eps^2\frac{\prt^2 u_m}{\prt x^2},\ldots\right)
\end{equation}
can be represented as a compatibility condition of two linear equations
\begin{equation}\label{A-3}
\begin{split}
\eps^2\chi_{xx}&=\mathcal{A}\chi, \\
\chi_t&=-\tfrac12\mathcal{B}_x\chi+\mathcal{B}\chi_x,
\end{split}
\end{equation}
where $\mathcal{A}$ and $\mathcal{B}$ depend on the $u_k$'s, their space
derivatives and on the spectral parameter $\la$. It is assumed that
the system (\ref{A-2}) has a periodic solution with wavelength
$L\propto\eps$ and it is parametrized by the constant parameters
$\la_i$ which appear in the finite-gap integration method in the
following way. The second-order linear equation (\ref{A-3}) has two
basis solutions $\chi_{\pm}$ and their product $g=\chi_+\chi_-$
satisfies a third-order differential equation which can be
integrated once to give
\begin{equation}\label{A-4}
\frac{\eps^2}{2} gg_{xx}-\frac{\eps^2}{4} g_x^2-\mathcal{A}g^2=
\sigma P(\la),
\end{equation}
where $\sigma$ is determined by the sign of the highest order term in
$\mathcal{A}$ as a function of $\la$,
i.e., $\mathcal{A}\sim-\sigma\la^r$ as $\la\to\infty$. Periodic
solutions are distinguished by the condition that $P(\la)$ is a
polynomial in $\la$ and then $\la_i$ are its zeros. We shall confine
ourselves to the one-phase periodic solutions which physical variables
depend on a single variable $x-V_{\varphi}\,t$ only.

In a modulated wave the parameters $\la_i$ become slow functions of
$x$ and $t$ whose evolution is described by the Whitham equations
which in the case (\ref{A-1}), (\ref{A-3}) can be written in the form
\begin{widetext}
\begin{equation}\label{A-5}
\frac{\prt \la_i}{\prt t}-
\frac{\langle\mathcal{B}/g\rangle}{\langle 1/g\rangle}\frac{\prt\la_i}{\prt x}
=\lim_{\eps\to0}
\left\{\frac{\sigma}{\langle 1/g\rangle\prod_{m\neq i}(\la_i-\la_m)}
\sum_k\left\langle\left(\frac{\prt\mathcal{A}}{\prt u_k}R_k
+\cdots+\frac{\prt\mathcal{A}}{\prt u_k^{(\ell_k)}}
\frac{\prt^{(\ell_k)}R_k}{\prt x^{(\ell_k)}}\right)g\right\rangle\right\},
\end{equation}
\end{widetext}
where $\ell_k$ denotes the highest order of derivative of $u_k$ entering
in $\mathcal{A}$. The angle brackets denote the averaging over
one wavelength:
\begin{equation}\label{A-6}
\left\langle\mathcal{F}\right\rangle=\frac1L\int_0^L\mathrm{d}x\,\mathcal{F}.
\end{equation}
The spectral parameter $\la$ should be put equal to $\la_i$ after averaging.

We shall apply here this scheme to the perturbed nonlinear
Schr\"odinger (NLS) equation
\begin{equation}\label{A-7}
{\rm i}\eps\,\psi_t+\tfrac12\eps^2\psi_{xx}-|\psi|^2\psi=
{\rm i}G(|\psi|^2)\psi,
\end{equation}
where $G(\rho)$ is a real function of the density $\rho=|\psi|^2$.
Eq.~\eqref{eq:4-1} pertains to this type [with $G(\rho)=\eta(1-\rho)$]. In
the case of Eq.~(\ref{A-7}) we have two field variables $\psi$,
$\psi^*$, and, correspondingly, two terms of perturbation in
(\ref{A-1}):
\begin{equation}\label{A-8}
R_{\psi}=G(\rho)\psi/\eps, \quad
R_{\psi^*}=G(\rho)\psi^*/\eps.
\end{equation}
For non-perturbed NLS equation the linear system (\ref{A-3}) is specified as
\begin{align}\label{A-9}
\mathcal{A}&=-\la^2-{\rm i}\eps\la\frac{\psi_x}{\psi}+\psi^*\psi-
\frac{\eps^2}2\frac{\psi_{xx}}{\psi}+
\frac{3\eps^2}4\frac{\psi_x^2}{\psi^2}, \\
\label{A-10}
\mathcal{B}&=-\la+\frac{{\rm i}\eps}2\frac{\psi_x}{\psi}.
\end{align}
Substitution of (\ref{A-9}) into (\ref{A-5}) shows that, in the
expression to be averaged [in the right-hand side of (\ref{A-5})], the
leading term in powers of $\eps$ is equal to $2\langle G\rho
g\rangle/\eps$. The averaging can be performed with the use of
equations known from the theory of periodic solutions of the NLS
equation (see, e.g., Ref.~\onlinecite{kamch2000}):
\begin{equation}\label{A-11}
\begin{array}{ll}
\vspace{3mm}
\displaystyle{g=\la-\mu_a,} &
\displaystyle{\eps\frac{\mathrm{d}\mu_a}{\mathrm{d}x}=2\sqrt{-P(\mu_a)},} \\
\vspace{3mm}
\displaystyle{-\frac{{\rm i}\eps}2\frac{\psi_x}{\psi}=\frac{s_1}2-\mu_a,} &
\displaystyle{V_{\varphi}=\frac{s_1}2,} \\
\displaystyle{L=\eps\oint\frac{\mathrm{d}\mu_a}{2\sqrt{-P(\mu_a)}},}
\end{array}
\end{equation}
where $P(\mu_a)=\prod_i(\mu_a-\la_i)$ and $s_1=\sum_i\la_i$. 
The quantity $\mu_a$ is
known as the auxiliary eigenvalue in the finite-gap integration method.
Hence, we obtain
\begin{equation}\label{A-12}
\begin{split}
& \left \langle \frac1g \right \rangle
= \left \langle \frac1{\la-\mu_a} \right \rangle
= - \frac2L \frac{\prt L}{\prt \la_i},
\\
& \left \langle \frac{\mathcal{B}}g \right \rangle
= - 1 + \frac{s_1}L \frac{\prt L}{\prt \la_i}.
\end{split}
\end{equation}
For calculating $\langle G\rho g\rangle$ we also take into account
that $\mu_a$ can be expressed as a function of $\rho$ in the following
way (see Ref.~\onlinecite{kamch2000}):
\begin{equation}\label{A-13}
\mu_a(\rho)=\frac{s_1}4+\frac{-j+{\rm i}\sqrt{\mathcal{R}(\rho)}}{2\rho},
\end{equation}
where
\begin{equation}\label{A-14}
\begin{split}
&\mathcal{R}(\nu)=(\nu-\nu_1)(\nu-\nu_2)(\nu-\nu_3), \\
&j^2=\nu_1\nu_2\nu_3,
\end{split}
\end{equation}
\begin{equation}\label{A-15}
\begin{split}
\nu_1=\tfrac14(\la_1-\la_2-\la_3+\la_4)^2,\\
\nu_2=\tfrac14(\la_1-\la_2+\la_3-\la_4)^2,\\
\nu_3=\tfrac14(\la_1+\la_2-\la_3-\la_4)^2,
\end{split}
\end{equation}
and
\begin{equation}\label{A-16}
\eps\frac{\mathrm{d}\rho}{\mathrm{d}x}=2\sqrt{\mathcal{R}}.
\end{equation}
Then we obtain the Whitham equations for the Riemann invariants
$\la_i$ in the form
\begin{equation}\label{A-17}
\begin{split}
\frac{\prt\la_i}{\prt t}+v_i\frac{\prt\la_i}{\prt x}
&=-\frac{v_i-s_1/2}{\prod_{m\neq i}(\la_i-\la_m)}\\
&\times\frac2L\int_{\nu_1}^{\nu_2}\mathrm{d}\nu\,
\frac{G(\nu)\big[(\la_i-s_1/4)\nu-j/2\big]}
{\sqrt{\mathcal{R}(\nu)}},
\end{split}
\end{equation}
with
\begin{equation}\label{A-18}
v_i=\frac{s_1}2+\left(\frac2L\frac{\prt L}{\prt\la_i}\right)^{-1}, \quad i\in\{1,2,3,4\}.
\end{equation}
In the stationary case, i.e., when $\prt\la_i/\prt t=0$ and
$s_1=2\,V_{\varphi}=0$, the Whitham equations simplify to
\begin{equation}\label{A-19}
\frac{\mathrm{d}\la_i}{\mathrm{d}x}=\frac2L \,
\frac{G_1 \la_i+G_2}{\prod_{m\neq i}(\la_i-\la_m)},
\end{equation}
where
\begin{equation}\label{A-20}
G_1=-\int_{\nu_1}^{\nu_2}\mathrm{d}\nu\,
\frac{\nu \, G(\nu)}{\sqrt{\mathcal{R}(\nu)}}, \quad
G_2=-\frac{j}{2}\int_{\nu_1}^{\nu_2}\mathrm{d}\nu\,
\frac{G(\nu)}{\sqrt{\mathcal{R}(\nu)}}.
\end{equation}
For $G(\rho)=\eta(1-\rho)$ we arrive at Eqs.~\eqref{eq:14-1} and \eqref{14-2}.


\begin{thebibliography}{99}
\bibitem{remcs} We use the convenient denomination ``speed of sound''
  for $c_s$ although very long-wavelength modes do not propagate in
  presence of damping and thus the sound velocity is only properly
  defined in absence of damping (see however the remark after
  Eq. \eqref{eq:qstar} in Sec.~\ref{3a}).
\bibitem{rem2} As discussed in the paragraph below
  Eq.~\eqref{eq:drag-large-V}, the drag force is even totally
  dominated by the wave resistance when $M\to\infty$.
\bibitem{rem3} This is because $\hat{f}_{\rm ext}(q_{\sss
    M}\to\infty)=0$. Roughly speaking, this means that the wave drag is
  negligible when the de Broglie wavelength $\hbar/(mV)$ gets smaller
  than the typical size of the potential.
\bibitem{remh} Note that the hydraulic approximation is also valid for
  describing the long-distance upstream wake in the subsonic case\cite{Kam12}.
\bibitem{rem} Note that $q_3$ initially slightly increases above $2{\rm i}$
  when $M\simeq 0$ [cf. Eq.~\eqref{eq:ap6}]. For legibility this
  small feature is not accounted for by the direction of the red arrow
  in Fig.~\ref{fig-app}.
\end{thebibliography}
\end{document}